\documentclass[acmtog]{acmart}
\acmSubmissionID{344}
\usepackage{booktabs, tabulary}
\setcopyright{rightsretained}
\acmJournal{TOG}
\acmYear{2021}
\acmVolume{40}
\acmNumber{6}
\acmArticle{253} 
\acmMonth{12}
\acmDOI{10.1145/3478513.3480551}

\newtheorem{theorem}{Theorem}[section]
\theoremstyle{definition}
\newtheorem{definition}{Definition}[section]

\usepackage{algorithmicx}
\usepackage{algorithm}
\usepackage{algpseudocode}

\DeclareMathOperator*{\argmin}{arg\,min}

\newcommand{\zoe}[1]{{\textbf{\scriptsize{\color{purple}Zoë: #1}}}}

\newcommand{\R}{\mathbb{R}}
\newcommand{\vol}{\text{Vol}}
\newcommand{\D}{\mathbb{D}}

\newcommand{\bfx}{\textbf{x}}
\newcommand{\bfv}{\textbf{v}}
\newcommand{\bfy}{\textbf{y}}
\newcommand{\bfa}{\textbf{a}}

\newcommand{\bfu}{\textbf{u}}
\newcommand{\Mo}{\mathcal{M}}
\newcommand{\bfs}{\textbf{s}}

\newcommand{\bff}{\textbf{f}}
\newcommand{\bfc}{\textbf{c}}
\newcommand{\bfp}{\textbf{p}}
\newcommand{\psdt}{\mathcal{S}_{++}^3}
\newcommand{\psdf}{\mathcal{S}_{++}^4}
\newcommand{\sos}{\Sigma}
\newcommand{\dcorr}{d_{\text{correct}}}
\newcommand{\dexact}{d_{\text{exact}}}
\newcommand{\mathcolorbox}[2]{\colorbox{#1}{$\displaystyle #2$}}

\usepackage{bm}
\usepackage{wrapfig}

\usepackage{listings}
\usepackage{color}
\usepackage{multicol}
\usepackage{vwcol}
\usepackage{courier}
\usepackage{parcolumns}
\usepackage{soul}
\usepackage{xcolor}
\newlength\dlf
\newcommand\alignedbox[3]{
  &
  \begingroup
  \settowidth\dlf{$\displaystyle #2$}
  \addtolength\dlf{\fboxsep+\fboxrule}
  \hspace{-\dlf}
  \fcolorbox{white}{#1}{$\displaystyle #2 #3$}
  \endgroup
}
\usepackage{wrapfig}
\usepackage{pgfplots}
\usepackage{pgfplotstable}
\usepackage{cleveref}

\newcommand{\PM}{\mathcal{P}}
\newcommand{\Exp}{\mathbf{E}}

\begin{document}

\title{Sum-of-Squares Geometry Processing}

\author{Zo\"e Marschner}
\email{zoem@mit.edu}
\author{Paul Zhang}
\email{pzpzpzp1@mit.edu}
\author{David Palmer}
\email{drp@mit.edu}
\author{Justin Solomon}
\email{jsolomon@mit.edu}
\affiliation{%
  \institution{
  \\
  Massachusetts Institute of Technology}
  \streetaddress{77 Massachusetts Avenue}
  \city{Cambridge}
  \state{MA}
  \postcode{02139}
  \country{USA}}

\authorsaddresses{Authors’ address: Zo\"e Marschner, zoem@mit.edu; Paul Zhang, pzpzpzp1@mit.edu; David Palmer, drp@mit.edu; Justin Solomon, jsolomon@mit.edu, Massachusetts Institute of Technology, 77 Massachusetts Avenue, Cambridge, MA, 02139, US}

\begin{abstract}
  Geometry processing presents a variety of difficult numerical problems, each seeming to require its own tailored solution. This breadth is largely due to the expansive list of \emph{geometric primitives}, e.g., splines, triangles, and hexahedra, joined with an ever-expanding variety of \emph{objectives} one might want to achieve with them. 
  With the recent increase in attention toward \emph{higher-order surfaces}, we can expect a variety of challenges porting existing solutions that work on triangle meshes to work on these more complex geometry types.
  In this paper, we present a framework for solving many core geometry processing problems on higher-order surfaces. We achieve this goal through sum-of-squares optimization, which transforms nonlinear polynomial optimization problems into sequences of convex problems whose complexity is captured by a single \emph{degree} parameter. This allows us to solve a suite of problems on higher-order surfaces, such as continuous collision detection and closest point queries on curved patches, with only minor changes between formulations and geometries.
\end{abstract}



\begin{teaserfigure}
  \includegraphics[width=\textwidth]{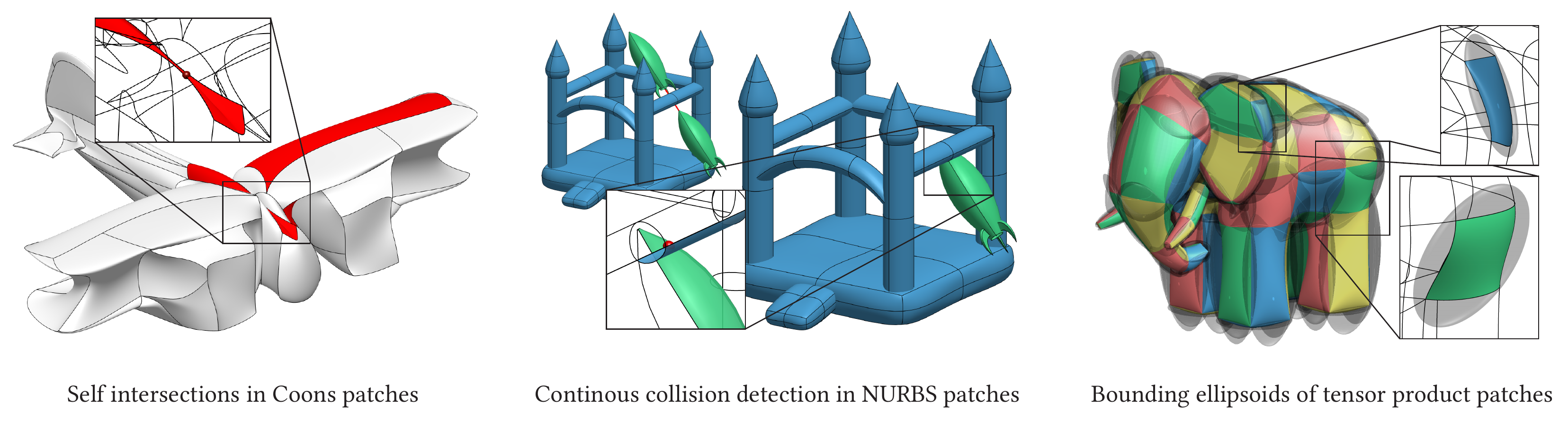}
  \vspace{-15pt}
  \caption{We describe a framework for solving many different geometry processing problems on polynomial patches using sum-of-squares relaxation. Here, we show three of the problems we discuss in this paper: Self-Intersection, Continuous Collision Detection, and Minimal Bounding Ellipsoid. Our formulation is very general, working for any piecewise polynomial or rational patch type. We show each problem here on a model with a different patch type: cubic Coons patches from \cite{smirnov2020learning}, bicubic NURBS patches from \cite{Trusty_Chen_Levin_SEM_2021}, and bicubic Bézier tensor patches. 
  }
  \label{fig:teaser}
\end{teaserfigure}

\maketitle
\section{Introduction}

Of the many geometric representations available today, \emph{polynomial patches} are exceptionally powerful. 
They are a mainstay of computer-aided design and digital sculpting, where they provide 
piecewise-smooth parametrization. They can represent complex shapes with just a few control points. And in finite-element modeling, high degree polynomial bases can be used to construct high-fidelity solutions to partial differential equations.


Polynomial patches have recently made inroads into geometry processing problems involving PDEs and simulation
\cite{Schneider:2019:PFM, Schneider:2018:DSA, mandad2020bezier,Trusty_Chen_Levin_SEM_2021,jiang2020highordertetmeshes}. However, basic geometric kernels that are straightforward to implement for piecewise-linear meshes remain challenging for higher-order patches. These kernels include detecting self-intersections and collisions, computing bounding volumes, and measuring distances. Such kernels become especially important in \emph{dynamics} problems, which in principle require continuous-time maintenance of physical feasibility.

We propose to bring the methods of \emph{sum-of-squares programming} (SOS) to bear on this domain. The core idea of SOS programming is the replacement of polynomial positivity constraints with more computationally tractable SOS constraints of bounded degree $d$, which can be represented by semidefinite programs. The miracle of SOS programming comes from the Positivstellensatz \cite{blekherman2012semidefinite}, which states that for a large enough $d$, the globally-optimal solution to the modified problem is certifiably equivalent to the globally-optimal solution of the original problem. This certificate of correctness comes from the theory of SOS programming and is known as \emph{exact recovery}. Thus, the SOS machinery allows us in many cases to solve seemingly non-convex optimization problems to global optimality. 

A caveat to SOS relaxation is that the cost of solving the relaxed problem increases factorially with $d$ \cite{blekherman2012semidefinite}. While the minimal required $d$ for exact recovery is problem-dependent and can be large in general, it is often close to the degrees of the polynomials in the original problem formulation, thereby maintaining tractability. Furthermore, we find that in practice the $d$ required to obtain the correct solution to the original problem is often lower than the $d$ required to obtain an exact recovery certificate. Since one often only cares about correctness of the solution, choosing the lower $d$ allows the cost of the problem to be driven down in this case. If an algebraic proof of correctness is desired, the higher $d$ can be chosen, at the expense of an increase in the cost of the problem.

Returning to geometry processing on polynomial patches, many geometric kernels can be formulated as polynomial optimizations. For example, surface-surface intersection (SSI) of quadratic triangles can be written as the minimization of a quartic objective function with linear inequality constraints.
Applying SOS relaxation then yields a convex problem directly, and it only remains to verify that the relaxation successfully produces a solution to the original problem, i.e., that $d$ was large enough. As illustrated by this example, the simplicity of SOS relaxation makes it readily adaptable to a wide variety of problems.

In this paper, we apply SOS methodology to several core problems in geometry processing on polynomial surfaces of varying degree. 
We show that a minor modification to our problem formulation allow us to support rational surfaces as well---namely, those consisting of NURBS patches.
We verify the success of our SOS formulations on an exhaustive suite of test data. An overview of these results is provided in \autoref{tbl:combined}, where we show the experimentally determined minimum degree $\dcorr$ required to solve each problem with 100\% accuracy.
Finally, we apply these geometric kernels on various higher-order meshes, demonstrating the extensibility of geometry processing methods on linear meshes to higher-order surfaces.
With these low-level operations out of the way, we pave the way for development of higher-level geometry processing techniques on higher-order surfaces.

\section{Related Work}

\paragraph{SOS programming}
Sum-of-squares (SOS) programming is a type of convex relaxation in which polynomial positivity constraints are replaced by SOS constraints. These can be transformed further into semidefinite optimization problems (SDP), which are solvable in polynomial time via interior-point methods \cite{alizadeh1995interior, nesterov-nemirovskii, boyd2004convex}.
Modeling frameworks such as \textsc{yalmip} \cite{Lofberg2004,prajna2002introducing} convert SOS formulations into SDPs, which can be solved by black-box solvers such as \textsc{mosek} and \textsc{sdpt3} \cite{aps_mosek_2017,tutuncu2001sdpt3}.
We provide the mathematical background relevant to this paper in \Cref{sec:prelim}. For a comprehensive review of this field, see \cite{blekherman2012semidefinite}.

\paragraph{SOS programming in geometry processing (GP)}
The use of SOS programming in geometry processing is fairly recent. Closest to our work is that of \citet{marschner2020hexahedral}, where SOS programming is applied to verify injectivity of trilinear hexahedra and to repair hexahedral meshes failing injectivity. We will show that this problem fits into our generalized framework. Prior to that, SOS programming was also used to find level set surfaces that encapsulate point clouds \cite{Ahmadi2017sos}.

\paragraph{Polynomial surfaces in GP} 
Polynomial surfaces have a long history in computer graphics and computer-aided design (CAD), with spline surfaces such as piecewise-bicubic Catmull-Clark surfaces \cite{catmull1978recursively} and piecewise-quartic Loop subdivision surfaces \cite{loop1987smooth} remaining popular in applications today. Recently geometry processing has seen a renaissance of polynomial meshes. \citet{mandad2020bezier} use B\'ezier curves to compute curvilinear triangle meshes of planar domains, while \citet{karvciauskas2020low} produce bicubic spline surfaces for locally quad-dominant meshes. \citet{jiang2020highordertetmeshes} develop a method to generate coarse higher-order tetrahedral meshes from linear ones.

Polynomial meshes have also proven useful in solving PDEs from geometry and simulation.
Catmull-Clark surfaces have been used for thin shell simulation \cite{PE:VMV:VMV11:113-120}. \citet{Schneider:2019:PFM, Schneider:2018:DSA} use higher-order basis functions to solve various PDEs, including linear elasticity. \citet{cardoze2004bezier} develop a method for maintaining quality of curved triangle meshes through fluid motion. 

\paragraph{GP problems for higher-order surfaces}
\citet{Trusty_Chen_Levin_SEM_2021} develop a method for simulation of elasticity on volumes enclosed by NURBS patches without volumetric remeshing. A notable gap in these methods is lack of exact collision detection and intersection prevention. 
We aim to show one path toward adding these key geometry kernels via SOS programming.

Surface-surface intersection (SSI) is the problem of finding intersection curves between surfaces. SSI is frequently used in CAD, where curves are obtained by first linearizing the surfaces to find an initial intersection point, followed by stepping along the common tangent direction of the two surfaces. This method requires tuning of various tolerance parameters, as well as a sufficiently dense initial linearization. We refer the reader to \cite{barnhill1987surface} for a summary of SSI in CAD. \citet{pekerman2008self} present several algorithms for self-intersection detection and removal in curves and surfaces. \citet{oh2012efficient} provide a culling approach combined with Newton-Raphson iterations to compute closest point projection to freeform curves and surfaces.

\section{Preliminaries}
\label{sec:prelim}

For a complete review of sum-of-squares theory we refer the reader to \cite{blekherman2012semidefinite,ParilloClass};
\citet{marschner2020hexahedral} also present similar background in the context of a geometry processing problem.
For completeness, we recall the most relevant concepts here. 

\subsection{Positive Polynomials and SOS Polynomials}
Let $\R[\bfu] = \R[u_1,...,u_k]$ be the ring of real multivariate polynomials in $\bfu$. $\R[\bfu]_d$ denotes the subset of polynomials of degree at most $d$. We will use $[\bfu]_d$ to denote the basis of monomials up to degree $d$. Any member $f(\bfu)\in\R[\bfu]_d$ can be written in this basis: $f(\bfu)=[\bfu]_d^\top \bff$, where $\bff$ denotes the vector of coefficients of monomials in $f$.

A special subset of $\R[\bfu]$ (resp.\ $\R[\bfu]_d$) is the cone of \emph{positive polynomials} $P$ (resp.\ $P_d$). As the name suggests, positive polynomials $f(\bfu)\in P$ satisfy $f(\bfu)\geq0$ for all $\bfu$. Many polynomial optimization problems can naturally be written with positive polynomial constraints, and thus it is highly desirable to be able to optimize over $P$. Unfortunately, even membership testing in $P$ is NP-hard in general \cite[\S 3.4.3]{blekherman2012semidefinite}. 

This leads us to the more restrictive subset $\Sigma$ of \emph{sum-of-squares (SOS) polynomials}; we use $\Sigma_d$ to denote SOS polynomials of bounded degree $d$. Members of this set $f(\bfu)\in\Sigma$ can be decomposed into sums of squares of polynomials: $f(\bfu)=\sum_i s_i(\bfu)^2$ for $s_i\in\R[\bfu]$. Naturally, they form a subset of $P$, giving us the following inclusions:
\begin{equation} \Sigma_d \subset P_d \subset \R[\bfu]_d \end{equation}
Unlike $P_d$, $\Sigma_d$ is computationally tractible---feasibility and optimization problems over $\Sigma_d$ translate naturally in to SDPs. This will be made explicit in \cref{sec:sosopt}.

\subsection{SOS Optimization}
\label{sec:sosopt}
Membership in $\Sigma_d$ can be expressed via semidefinite programming.
Using $\langle\cdot,\cdot\rangle$ to denote the Frobenius inner product, an SOS polynomial is equivalently written as
\begin{equation} \begin{aligned} f(\bfu) &= \begin{pmatrix} s_1(\bfu) \\ \vdots \\ s_k(\bfu) \end{pmatrix}^\top \begin{pmatrix} s_1(\bfu) \\ \vdots \\ s_k(\bfu) \end{pmatrix} \\
&= [\bfu]_{\lfloor d/2 \rfloor}^\top 
\underbrace{
\begin{pmatrix} \bfs_1 & \cdots & \bfs_k \end{pmatrix}
\begin{pmatrix} \bfs_1^\top \\ \vdots \\ \bfs_k^\top \end{pmatrix} 
}_{S}
[\bfu]_{\lfloor d/2 \rfloor} \\
&= \left\langle S, [\bfu]_{\lfloor d/2 \rfloor}[\bfu]_{\lfloor d/2 \rfloor}^\top \right\rangle,
\end{aligned} \label{eq:SOSSDP}\end{equation}
\noindent where the coefficients $\bfs_i$ are now encoded in the matrix $S$. \eqref{eq:SOSSDP} provides a linear relationship between $\bff$ and $S$, which is positive semidefinite by construction. Indeed, we have shown that the existence of such an $S \succeq 0$ is equivalent to $f \in \Sigma_d$. See \Cref{tbl:combined} for rough dimensions of the SDP problem.

Using the above transformation to render SOS constraints into semidefinite constraints, one can perform optimization over $f(\bfu)\in\Sigma_{d}$ with relative ease. For this reason, it is often profitable to transform a problem involving a positivity constraint $f\in P$ to one with a set of constraints of the form $s_i\in\Sigma_d$---resulting in an \emph{SOS program}.
A key theorem in SOS programming, the \emph{Positivstellensatz}, explains how to effect this transformation so that the global optimum of the original problem is recovered in the limit of increasing SOS degree $d$ \cite{blekherman2012semidefinite}. We will use a specialized version of the Positivstellensatz below.

\subsection{SOS Optimization on a Compact Domain}
\label{sec:SOS_compact}
In the context of geometry processing, one frequently seeks to optimize a functional over a compact domain, rather than over all of $\R^k$. Thus, one often encounters constraints of the form $f(\bfu) \ge 0$ for $\bfu \in \D$ where $\D$ is compact. SOS programming can be extended to handle such constraints. Here the key theorem is Putinar's variant of the Positivstellensatz:
\begin{theorem}[Putinar's Positivstellensatz \cite{putinar1993positive}; see also \cite{blekherman2012semidefinite}, Theorem 3.138] \label{thm:putinar}
Let $\D = \{\bfu\in\R^k: g_i(\bfu)\geq0\}$ be a domain with an algebraic certificate of compactness. Any polynomial $f(\bfu)$ that is strictly positive on $\D$ admits a decomposition
\begin{equation}
f(\bfu) = s_0(\bfu) + \sum_{i = 1}^m s_i(\bfu) g_i(\bfu), \label{eq:putinar}
\end{equation}
with SOS polynomials $s_i \in \Sigma_d$ for high enough degree $d$.
\label{thm:putinar}
\end{theorem}
Observe that the decomposition \eqref{eq:putinar} provides a certificate of nonnegativity by construction, since all $s_i$ are nonnegative and all $g_i$ are nonnegative on $\D$.
For brevity we omit details of the required algebraic certificate of compactness and refer the interested reader to \cite{blekherman2012semidefinite}. 
For all domains encountered in this paper, compactness certificates are readily computable.
The theory of SOS programming has an elegant dual formulation that allows us to determine a sufficiently large $d$, but we will defer discussion of this dual to \Cref{sec:momentrelaxation} to prioritize presenting concrete examples of what SOS programming can achieve for geometry processing.
\begin{figure}
    \centering
    \includegraphics[width=1.0\columnwidth]{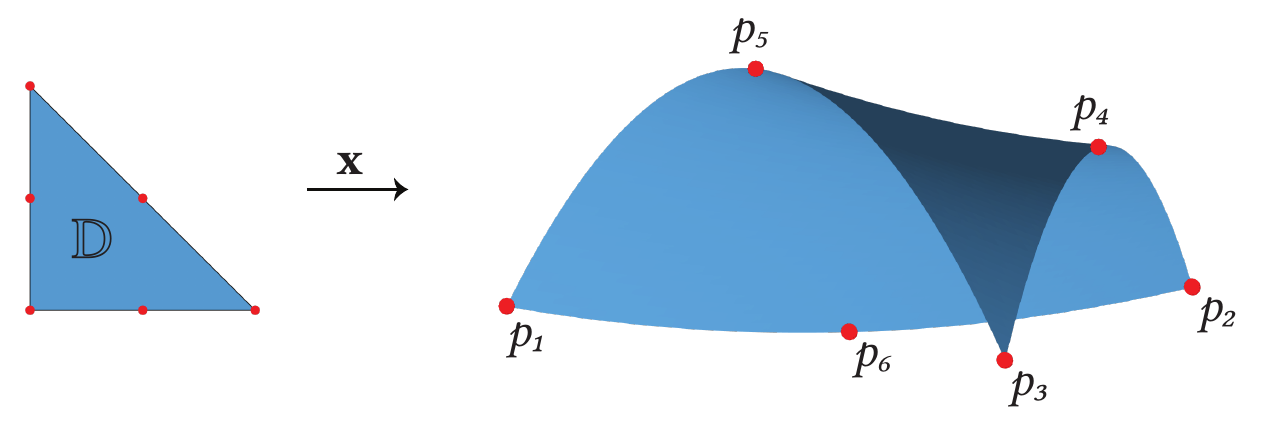}
    \vspace{-15pt}
    \caption{Quadratic triangle patch}
    \label{fig:quadrtri}
\end{figure}
\paragraph{Equality Constraints}

It is often convenient to consider \emph{semialgebraic} domains $\D$ defined by polynomial inequalities \emph{and equations}. This generalization does not pose a significant problem for SOS methods, for the reasons we will outline below.

Given a domain defined by equations and inequalities $\D = \{\bfu : g_i(\bfu) \ge 0, h_i(\bfu) = 0\}$, a decomposition of the form
\begin{equation}
    f(\bfu) = s_0(\bfu) + \sum_{i=1}^m s_i(\bfu) g_i(\bfu) + \sum_{i=1}^n p_i(\bfu) h_i(\bfu), \label{eq:equality-certificate}
\end{equation}
where the $s_i$ are SOS and the $p_i$ are arbitrary polynomials,
certifies the nonnegativity of $f$ on $\D$.

One can formally rewrite each equality constraint $h_i(\bfu) = 0$ as a pair of inequality constraints $h_i(\bfu) \ge 0$ and $-h_i(\bfu) \ge 0$, reducing $\D$ to the form required in \Cref{thm:putinar}. Then the theorem shows existence of a certificate of the form
\begin{equation}
    f(\bfu) = s_0(\bfu) + \sum_{i=1}^m s_i(\bfu) g_i(\bfu) + \sum_{i=1}^n (t_i^+(\bfu) - t_i^-(\bfu)) h_i(\bfu),
\end{equation}
where the $s_i$, $t_i^+$, and $t_i^-$ are SOS. But such a certificate is \emph{a fortiori} of the form \eqref{eq:equality-certificate}.
In this way, \Cref{thm:putinar} extends to the case of mixed constraints.

%

\subsection{Polynomial Patches}
\begin{definition}
A \emph{polynomial patch} is a map $\mathbf{x}:\mathbb{D} \subset \R^k \rightarrow \R^n$ from a compact semialgebraic base domain $\mathbb{D} = \{\textbf{u}\in\R^k : g_i(\textbf{u})\geq 0\}$, and such that each component of $\mathbf{x}$ is a multivariate polynomial of bounded degree $d_{\bfx}$.
\end{definition}
In practice, the base domain is typically a canonical triangle, square, cube, or similar, and so the polynomials $g_i$ are affine.
The dimension of the patch for most geometry processing tasks is $k \in \{1,2,3\}$, and $n$ controls the embedding dimension of the patch. 

It is often useful to parameterize these polynomials $\mathbf{x}$ by linear combinations of a few basis functions, with coefficients that are obtained from control points $\bfp_i \in \R^n$. Patches of this form include B\'ezier and Coons patches, and they generalize the simplest polynomial patch, the \emph{linear triangle}, which we will describe here for illustrative purposes. For the linear triangle patch, $\mathbb{D}$ is the triangle with vertices $(1,0),(0,1),(0,0)$, $n=3$, and the basis functions (in barycentric coordinates) are $\phi_1(\bfu) = u_1$, $\phi_2(\bfu) = u_2$, and $\phi_3(\bfu) = 1 - u_1 - u_2$. If the vertices of the linear triangle in $\R^3$ are $\bfp_1$, $\bfp_2$ and $\bfp_3$, then 
$\bfx(\bfu) = \phi_1\bfp_1  + \phi_2\bfp_2  + \phi_3\bfp_3$.

By varying the degree and number of basis functions, one can realize a variety of patch types. For example, a quadratic triangle is parameterized by six basis functions with control points $p$ located at the vertices and edge midpoints. Such a quadratic triangle, along with its control points, is depicted in \Cref{fig:quadrtri}. We will use $n_B$ to denote the number of basis functions needed for a particular patch. For the purposes of this paper, we will focus our attention on quadratic and cubic triangles, quadratic Bézier curves, bicubic Bézier tensor patches, cubic Coons patches, and B-spline surfaces.

\section{Geometric Kernel Problems}
\label{sec:problems}
In this paper, we show that a variety of geometric problems on polynomial patches can be solved with SOS programming. Our key observation is that most of these can be formulated in a very similar way. In particular, we offer the following template problem:
\begin{equation}\label{eq:templateproblem}
\begin{alignedat}{2}
f^* &= \underset{\bfu \in \D}{\text{min}} \; && f(\textbf{u}) \\
\bfu^* &= \argmin_{\bfu \in \D} \; && f(\bfu).
\end{alignedat}
\end{equation}
where
\begin{equation} \mathbb{D} = \{\textbf{u}\in\R^k : g_i(\textbf{u})\geq 0, h_i(\textbf{u}) = 0\}, \end{equation}
is a compact semialgebraic domain.
In what follows, we will outline how to compute the globally optimal value $f^*$ using the machinery of SOS programming, and we will apply this to a variety of patch problems. In \Cref{sec:momentrelaxation}, we will discuss how to compute $\bfu^*$ and determine the correct degree for the relaxation.

To apply SOS methods, we rewrite the problem in terms of an additional variable $\lambda$, which acts as a lower bound on $f(\bfu)$ for $\bfu\in\D$. This is equivalent to asking that the polynomial $f(\bfu)-\lambda$ be positive for $\bfu\in\D$.
Applying \Cref{thm:putinar}, such a requirement is equivalent to the SOS constraints
\begin{equation} f-\lambda-\sum_i h_i p_i - \sum_i g_i s_i \in \sos, \quad s_i \in \sos, \quad p_i \in \R[\bfu]. \end{equation}
We have thus relaxed the problem \eqref{eq:templateproblem} into the SOS form
\begin{equation}\label{eq:templateproblemsos}
    \begin{aligned}
        \lambda^* &= \left\{\begin{aligned}
       & \max_{\lambda \in \R} \quad \lambda &&\\
       & \mathrm{s.t.} \quad\quad f - \lambda - \sum_i h_i p_i - \sum_i g_i s_i \in \sos_d, &\quad& \\
       & \mathrm{} \,\quad\quad\quad s_i \in \sos_d \\
       & \mathrm{} \,\quad\quad\quad p_i \in \R[\bfu]_d
        \end{aligned}\right\}.
    \end{aligned}
\end{equation}
As $d$ increases, \Cref{thm:putinar} guarantees that $\lambda^*$ converges to the globally optimal value $f^*$. For high-enough $d$, the relaxation will be accurate to within numerical precision, and as we observe in \Cref{sec:results}, the convergence is dramatic at a fairly low $d$.

\subsection{Optimization over a Polynomial Patch} \label{sec:singlepatch}
The template problem \eqref{eq:templateproblem} allows us to optimize arbitrary polynomial objective functions over a polynomial patch. One only needs to choose the $g_i$ that encode a base domain $\mathbb{D}$ along with an appropriate objective function $f$. We now give several examples of objective functions of interest to geometry processing.

\paragraph{Closest Point (CP)}
The closest point problem aims to find the minimum distance between a target point $\textbf{t}$ and a polynomial patch. Let $\bfx(\bfu)$ be the shape function of the patch. Then one simply chooses the objective function $f(\textbf{u}) = \|\textbf{x}(\textbf{u}) - \textbf{t}\|_2^2$ in the template problem.

\paragraph{Minimal Axis Aligned Bounding Box (MBB)}
The axis aligned bounding box problem finds the smallest-volume AABB containing a polynomial patch $\bfx(\bfu)$. The bounds can be obtained by choosing the objective function $f(\textbf{u}) = \pm x_i(\textbf{u})$. 

\paragraph{Hexahedron Quality Evaluation}
The hexahedral quality evaluation problem presented by \citet{marschner2020hexahedral} also fits into the template. Here $\D$ is a unit cube, $\bfx(\bfu)$ encodes trilinear interpolation, and $f(\bfu)$ is the Jacobian determinant of $\bfx(\bfu)$, i.e., $f(\bfu)=\det(\nabla_{\bfu} \bfx)$.

\subsection{Multiple Patches in One Optimization}
\label{sec:multipatch}
Often it is desirable to optimize an objective over multiple patches, or multiple copies of the same patch, simultaneously. Such problems can often be recast as optimization problems over a higher-dimensional base domain formed from the Cartesian product of the original base domains.

To wit, let polynomial patches $\textbf{x}^1$, and $\textbf{x}^2$ map respectively from base domains $\mathbb{D}_1 = \{\bfu^1 : g^1_i(\bfu^1) \ge 0\}$, and $\mathbb{D}_2 = \{\bfu^2 : g^2_i(\bfu^2) \ge 0\}$ into $\R^n$.
Let
\begin{equation} \begin{aligned} \mathbb{D} &= \{ (\bfu^1, \bfu^2) : g^1_i(\bfu^1) \ge 0, g^2_i(\bfu^2) \ge 0, q_j(\bfu^1, \bfu^2) \ge 0 \} \\ &\subseteq \mathbb{D}_1 \times \mathbb{D}_2, \end{aligned} \end{equation}
where the $q_j$ are additional problem-dependent and/or symmetry-breaking constraints.
One can then define a new patch $\textbf{x}(\bfu^1, \bfu^2) = (\textbf{x}^1(\bfu^1), \textbf{x}^2(\bfu^2))$ and rewrite $f$ relative to $\textbf{x}$ and $\D$. For notational convenience, we will continue to use $\textbf{u}^1$, $\textbf{u}^2$, $\textbf{x}^1$, and $\textbf{x}^2$ directly.


\paragraph{Patch Diameter (PD)}

\setlength{\intextsep}{2pt}
\setlength{\columnsep}{6pt}
\begin{wrapfigure}[5]{i}{0.25\columnwidth}
\includegraphics[width=0.25\columnwidth]{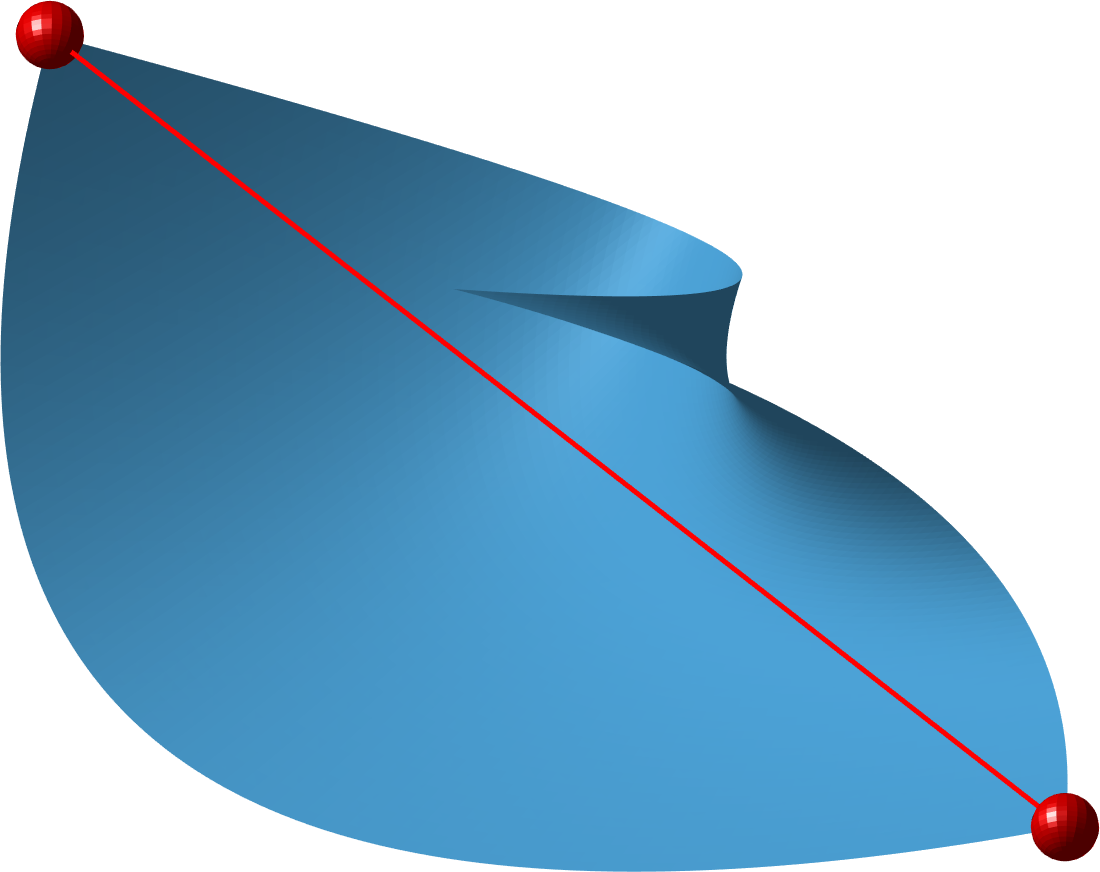}
\end{wrapfigure}

The \emph{diameter} of a patch is the distance between the two most distant points on that patch. This can be found using the template problem with two identical base domains $\mathbb{D}_1=\mathbb{D}_2$ and maps $\textbf{x}^1 = \textbf{x}^2$. Then one simply chooses the objective $f(\textbf{u}^1, \textbf{u}^2) = -\| \textbf{x}^1(\textbf{u}^1) - \textbf{x}^2 (\textbf{u}^2)\|_2^2$.

\paragraph{Surface-Surface Intersection (SSI)}
This problem detects when two polynomial patches intersect each other. As described previously, we use an augmented constraint set consisting of constraints from both patches. We further add the equality constraints $h(\textbf{u}^1, \textbf{u}^2) = \bfx^1(\bfu^1) - \bfx^2(\bfu^2) = 0$. This restricts optimization from the domain $\D_1 \times \D_2$ to its subset on which intersections occur. Feasibility of this problem determines if the two surfaces intersect. 
We can arbitrarily choose as an objective function $f(\textbf{u}^1, \textbf{u}^2) = u^1_1$.

\paragraph{Self-Intersection (SI)}
The self-intersection problem finds places where a single polynomial patch intersects itself. To fit this into the template problem, we start with the constraints of the SSI problem for two identical domains $\D_1 = \D_2$, $\bfx^1=\bfx^2$. Then we choose the objective to be $f(\textbf{u}^1, \textbf{u}^2) = - \| \textbf{u}^1 - \textbf{u}^2 \|_2^2$. This gives us points that are as far apart as possible in the base domain but map to the same embedded point in $\R^n$.

\begin{figure}
    \centering
    \includegraphics[width=1\columnwidth]{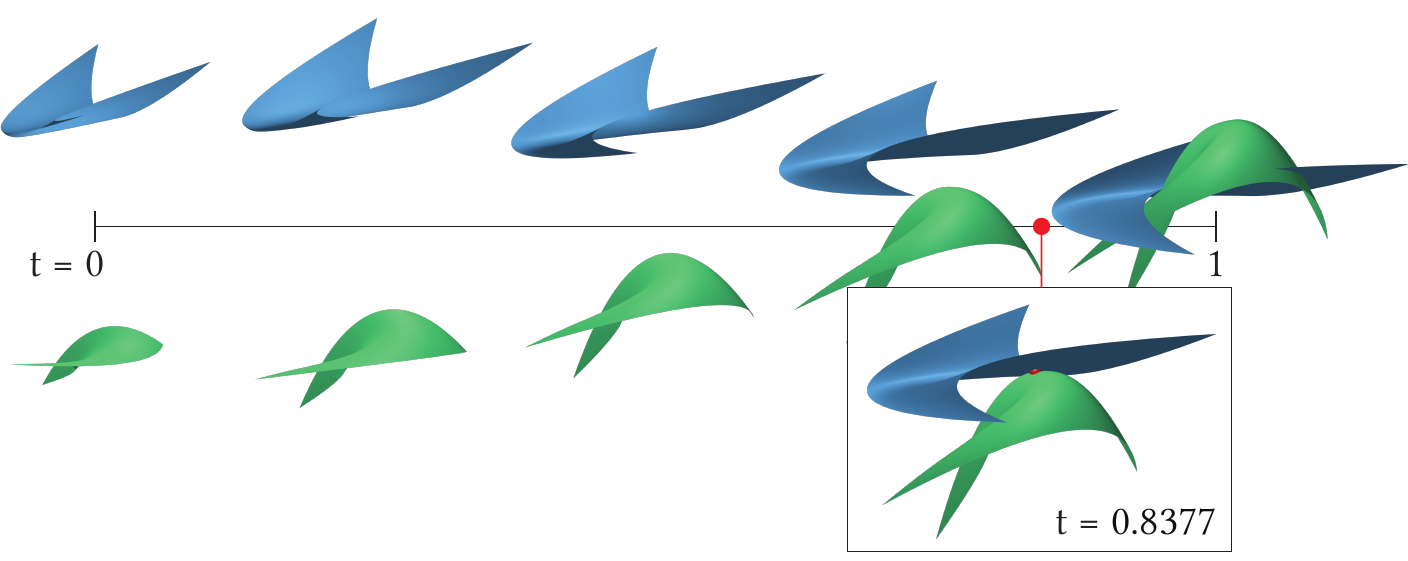}
    \vspace{-5pt}
    \caption{The CCD problem is demonstrated on two time-varying quadratic triangle patches over the time interval $[0, 1]$. The patches are shown at the time of first collision, at $t=0.8377$, in the inset, with the collision point marked in red.}
    \label{fig:CCD_ex}
\end{figure}

\paragraph{Continuous Collision Detection (CCD)}
The Continuous Collision Detection problem aims to find out if and when two patches on a time varying trajectory will intersect in a given time interval.
\Cref{fig:CCD_ex} shows an example of the CCD problem.
Let two patches $\bfx^1$, $\bfx^2$ be defined by control points $\bfp_i^1$, $\bfp_i^2$ via $\bfx(\bfu) = \sum_i^{n_B} \bfp_i \phi_i$. Let control points have velocities $v_i^1$ and $v_i^2$. We then define time-dependent patches by linear interpolation $\bfx(\bfu, t) = \sum_i^{n_B} (\bfp_i + \bfv_i t) \phi_i$, mapping from the augmented base domain $\D_1 \times \D_2\times[0, t_{max}]$ to $\R^n$. We wish to determine if the provided velocities will result in a collision within $t_{max}$ units of time.

As in the SSI case, we augment the constraint set with an equality constraint $h(\bfu^1,\bfu^2,t)= \bfx^1(\bfu^1, t) - \bfx^2(\bfu^2, t) = 0$. This restricts the optimization to the subset of the joint base domain on which space-time collisions occur. It just remains to choose $f(\bfu^1,\bfu^2,t) = t$ to find the earliest collision.

\subsection{Outside the Template} \label{sec:outside-template}
\paragraph{Rational Surfaces}
A natural generalization of polynomial surfaces are \emph{rational surfaces}---surfaces whose shape functions $\bfx$ are rational. Using control points $\bfp_i$ and basis functions $\phi_i$, rational surfaces are decomposable as follows 
\begin{equation}
    x_j(\bfu) = \frac{\sum_i p_{ij} \phi_i(\bfu)}{b(\bfu)} = \frac{a_j(\bfu)}{b(\bfu)},
\end{equation}
for polynomials $a_j,b,\phi_i\in\R[\bfu]$. Straightforward application of problem formulations from \Cref{sec:singlepatch} and \Cref{sec:multipatch} to rational surfaces may seem not to fit the template \eqref{eq:templateproblem}. The key observation for applying SOS programming to rational surfaces is that we simply need to clear the denominators. We demonstrate here the SSI problem on rational surfaces, with all other problems following \emph{mutatis mutandis}. The SSI formulation applied to rational surfaces starts with
\begin{equation}
\begin{aligned}
\label{eq:ssi_rational}
\begin{aligned}
& \underset{
\bfu^1 \in \D_1,  
\,\bfu^2 \in \D_2
}{\text{min}} & & u^1_1 \\
& \text{s.t.} & & 
\frac{\bfa^1(\bfu^1)}{b^1(\bfu^1)} = \frac{\bfa^2(\bfu^2)}{b^2(\bfu^2)} 
\end{aligned}
\end{aligned}
\end{equation}
By cross-multiplying denominators, the rational constraints are transformed back into polynomial constraints. 
\begin{equation}
\begin{aligned}
\label{eq:ssi_rational_homogenized}
\begin{aligned}
& \underset{
\bfu^1 \in \D_1,  
\,\bfu^2 \in \D_2
}{\text{min}} & & u^1_1 \\
& \text{s.t.} & & 
\bfa^1(\bfu^1)\,b^2(\bfu^2) = \bfa^2(\bfu^2)\,b^1(\bfu^1)
\end{aligned}
\end{aligned}
\end{equation}
Thus problems that can be solved on polynomial patches via SOS programming can be extended to rational surfaces. 

\paragraph{Minimal Enclosing Ellipsoid (MEE)}
The Minimal Enclosing Ellipsoid problem finds the ellipsoid of smallest volume that fully contains a polynomial patch. Smallest ellipsoids enclosing point sets have a long history in convex optimization, where they are known as \emph{L\"owner-John ellipsoids} \cite{todd2016minimum}. The MEE of a finite collection of points or ellipsoids can be computed exactly via SDP. This involves a reduction from the problem of computing the smallest ellipsoid with arbitrary center to the case of an ellipsoid centered at the origin. We will combine this reduction with SOS methods to compute the MEE of a polynomial patch.

We can parameterize an ellipsoid in $\R^3$ by a positive definite matrix $A \in \psdt$ and center point $\bfc\in \R^3$:
\begin{equation} \begin{aligned} \mathcal{E}(A,\bfc) &= \{ \bfp \in \R^n : (\bfp - \bfc)^\top A (\bfp - \bfc) \leq 1\}, \\
\vol(\mathcal{E}(A,\bfc)) &\propto \det(A^{-1}) = (\det A)^{-1}. \end{aligned} \end{equation}
Given $\D$ and $\bfx$ describing a 3D polynomial patch, the MEE can be computed as the solution to
\begin{equation}
\begin{aligned}
\label{eq:meefull}
(A^*, \bfc^*) = \left\{
\begin{aligned}
& \underset{A\in\psdt,\; \bfc \in \R^3}{\text{argmin}} & & \vol(\mathcal{E}(A,\bfc))\\
& \text{s.t.} & & 
1 \geq \underset{\textbf{u} \in \D}{\text{max}}\; (\bfx(\bfu) - \bfc)^\top A (\bfx(\bfu) - \bfc) 
\end{aligned}\right\}
\end{aligned}
\end{equation}
The problem in this form does not clearly fit into the template due to the inner maximization, which can also be viewed as a universal quantification. But as we will now show, we can still apply SOS optimization to obtain the MEE. The first step is to reduce the general MEE problem \eqref{eq:meefull} to an equivalent \emph{centered} MEE problem in one higher dimension:
\begin{equation}
\label{eq:cmeefull}
B^* = \left\{
\begin{aligned}
& \underset{B\in\psdf}{\text{argmin}} & & \vol(\mathcal{E}(B, 0))\\
& \text{s.t.} & & 1 \geq \underset{\textbf{u} \in \D}{\text{max}}\; \bfy^\top(\bfu) B \bfy(\bfu) \\
& & & \bfy(\bfu) = \begin{bmatrix}\bfx(\bfu)\\ 1\end{bmatrix}
\end{aligned}\right\}
\end{equation}
From $B^*$, one can recover $A^*$ by completing the square \cite{todd2016minimum}.
The constraints of \eqref{eq:cmeefull} require that the polynomial $1-\bfy^\top B \bfy \in \R[\bfu]$ be nonnegative for all $\bfu \in \D$. \Cref{thm:putinar} allows us to encode this requirement with SOS constraints, resulting in the following SOS program.
\begin{equation}
\label{eq:cmee_quadmod}
B^* = \left\{
\begin{aligned}
& \argmin & & \log(\vol(\mathcal{E}(B, 0))) \\[0.5em]
& \text{s.t.} & & s_i \in \sos, \;\;\;\;\; s_i \in \R[\bfu]_d\\
& & & 1 - \bfy^\top B \bfy - \sum_i g_i s_i \in \sos\\[-0.5em]
& & & B \in \psdf
\end{aligned} \right\},
\end{equation}
where we have also convexified the objective by taking its log. Both the objective and constraints are now convex, and we can solve \eqref{eq:cmee_quadmod} to global optimality, and by extension the MEE problem. 

Our ability to solve the MEE problem hinges on the fact that MEE has an equivalent centered formulation wherein $\bfc=0$. Consider what would happen if we tried to relax \eqref{eq:meefull} directly. 
We would require $1-(\bfx-\bfc)^\top A(\bfx-\bfc) \geq 0$ for $\bfu\in\D$. Our constraints would contain bilinear terms like $\bfx^\top A \bfc$ and linear-quadratic terms like $\bfc^\top A \bfc$, thus breaking convexity. While our solution to the MEE problem does not generalize directly to other problems outside the template, it illustrates what can and cannot be handled by SOS programming. 

\paragraph{Minimal Surrounding Sphere (MSS)}
The Minimal Surrounding Sphere problem finds the sphere of smallest volume that fully contains a polynomial patch. This can be solved with a minor modification to \eqref{eq:cmee_quadmod}. Let $B_3$ be the top left $3\times 3$ block of $B$. We simply add the constraint that $B_3 = a I_3$, where $I_3$ is the $3\times 3$ identity matrix and $a$ is a scalar variable. This additional constraint reduces the MEE program to an MSS program without affecting convexity.

\section{Moment Relaxation in Theory and Practice}
\label{sec:momentrelaxation}
While the previous section lists a variety of problems that can be solved with SOS programming, two points remain. First, while the SOS approach of the previous section gives us a way to compute the optimal value of \eqref{eq:templateproblem} by way of \eqref{eq:templateproblemsos}, we do not have a way to compute the $\argmin$ of \eqref{eq:templateproblem}. Using CP as an example, we can find the distance from a target point to its projection on a patch, but do not have the projected point. Second, we have not mentioned how to certify correctness of a solution and determine if $d$ is large enough for \Cref{thm:putinar} to hold. This section will address both points.

\begin{figure}
    \pgfplotstableread[col sep=comma]{figures/exact-vs-correct.csv}{\exactvscorr}
    \pgfplotstableread[col sep=comma]{figures/worst-xdists.csv}{\worstxdist}
    \begin{tikzpicture}
    \begin{axis}[
        width = \columnwidth,
        height = 0.8\columnwidth,
        enlarge x limits = 0,
        xlabel = {SOS Degree $d$},
        ylabel = {Percentage},
        ylabel near ticks,
        ytick pos = left,
        xticklabels = {},
        every tick label/.append style = {font=\tiny}]
    \addplot +[mark=none, blue, thick] table [x expr=\coordindex+1, y expr=100*\thisrowno{0}] {\exactvscorr};
    \addplot +[mark=none, red, thick] table [x expr=\coordindex+1, y expr=100*\thisrowno{1}] {\exactvscorr};
    \end{axis}
    \begin{axis}[
        axis y line* = right,
        width = \columnwidth,
        height = 0.8\columnwidth,
        enlarge x limits = 0,
        ylabel = {$\bfx$ Distance},
        ylabel near ticks,
        ytick pos = right,
        every tick label/.append style = {font=\tiny},
        legend style={at={(0.5,1.15)},
		anchor=north,legend columns=-1}
    ]
    \addplot+[mark=none,blue,thick] coordinates {(1, 0) (1, 0)};
    \addlegendentry{\% misidentified};
    \addplot+[mark=none,red,thick] coordinates {(1, 0) (1, 0)};
    \addlegendentry{\% exact};
    \addplot +[mark=none, orange, thick] table [x expr=\coordindex+1, y index=0] {\worstxdist};
    \addlegendentry{Worst $\bfx$ Distance};
    \end{axis}
    \end{tikzpicture}
\caption{For the SSI problem on quadratic triangles, all intersections are correctly identified for degree $d \ge 2$, and the distance between extracted points of intersection drops to 6.7e-05 at degree $d=5$, even though exact recovery does not yet occur at these degrees. Triangle control points were sampled from i.i.d. standard normal distributions.}
\label{fig:degrading}
\end{figure}
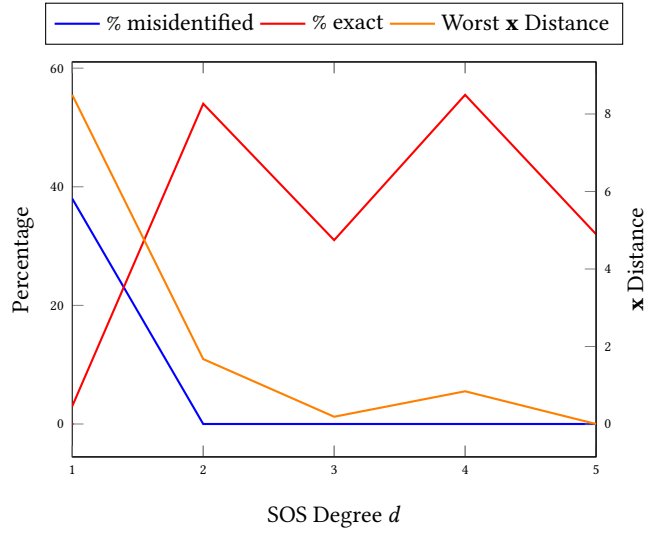

\subsection{Moment Relaxation in Theory}
Consider an alternative formulation of \eqref{eq:templateproblem} as a \emph{measure relaxation}: \begin{equation}
f^* \coloneqq \min_{\mu \in \PM(\D)} \Exp_\mu[f],
\label{eq:measureopt}
\end{equation}
where $\PM(\D)$ is the space of probability measures on $\D$, and $\Exp_\mu[f]=\int_{\D} f\,d\mu$ denotes integration against $\mu$. If $\bfu^* = \argmin_{\bfu \in \D} f(\bfu)$, then the optimum of \eqref{eq:measureopt} is achieved by placing all the mass of $\mu$ at $\bfu^*$---that is, the delta measure $\delta_{\bfu^*}$ minimizes \eqref{eq:measureopt} \cite{lasserre2001global}. While this relaxation convexifies \eqref{eq:templateproblem}, optimization over the infinite-dimensional space of measures $\PM(\D)$ is intractable.
To realize the measure relaxation computationally, $\mu$ must be represented by its moments. Using the multi-index $\bm{\alpha}$ to index monomial exponents of $f$, the $\bm{\alpha}$-moment of $\mu$ is defined as
$\mu_{\bm{\alpha}} \coloneqq \Exp_{\mu} [\bfu^{\bm{\alpha}}]$. Let $f_{\bm{\alpha}}$ be the coefficient of $\bfu^{\bm{\alpha}}$. Then the expectation in \eqref{eq:measureopt} can be expressed as
\begin{equation}
    \Exp_\mu[f] = \sum_{\bm{\alpha}} f_{\bm{\alpha}} \mu_{\bm{\alpha}}, \label{eq:pseudoexp}
\end{equation}
a simple linear function of \emph{finitely many} moments. It remains to constrain $\mu_{\bm{\alpha}}$ to be valid moments of a distribution over $\D$, a task fulfilled by the following theorem:
\begin{theorem}[\cite{lasserre2001global}, Theorem 4.2] \label{thm:lasserre}
Suppose $\D$ is a compact domain with algebraic certificate of compactness. 
Let $d$ be large enough for $f(\bfu) - f^*$ to admit the decomposition in \eqref{eq:putinar}. Then the following \textbf{moment relaxation} SDP computes $f^*$:
\begin{equation}
f^* = \left\{
    \begin{alignedat}{3}
  & \min_{\bm{\mu} \in \Mo_{d}} & & \Exp_{\bm{\mu}}[f(\bfu)] \\
   & \mathrm{s.t.}& \quad &\Exp_{\bm{\mu}}[q(\bfu)^2] \geq 0, \quad &&\forall q \in \R[\bfu]_{\lceil d/2\rceil} \\
   & & &\Exp_{\bm{\mu}}[q(\bfu)^2g_i(\bfu)] \geq 0, \quad &&\forall q \in \R[\bfu]_{\lceil d/2\rceil - w_i} \\
   & & &\Exp_{\bm{\mu}}[1] = 1,
    \end{alignedat}
    \right\}
\label{eq:lasserre}
\end{equation}
where $w_i = \lceil \deg g_i / 2 \rceil$, and $\Mo_d$ is the space of moment vectors up to degree $d$.
Furthermore, the moment vector corresponding to $\delta_{\bfu^*}$ minimizes \eqref{eq:lasserre}.
\end{theorem}
We now have two SDP-based approaches to solving the same polynomial optimization problem---one derived by SOS programming, and one via moment relaxation. As the reader may have guessed, the two SDPs are dual to each other \cite{lasserre2001global}.

\begin{table*}
    \centering
    \caption{Overview of SOS results on quadratic and cubic triangle patches.
    For each problem, we randomly generate 1000 test cases. Specifically for CCD on cubic triangles we test on a reduced set of 300 cases. We show the minimum SOS polynomial degree $\dcorr$ for which all problem instances are solved correctly. This degree is generally less than the degree $\dexact$ required for exact recovery. For problems where it is applicable (SSI,SI,CCD), we provide the percentage of the data that intersects or collides. We also provide the number of optimization variables for each problem, which indicate the dimensions of the ultimately solved SDP problem. Additionally, we provide the median runtimes over 50 instances of each problem with standard deviations. We separate out the time spent in \textsc{yalmip} converting to \textsc{mosek}'s input form, and the time spent in \textsc{mosek} actually solving the SDP problem. CCD on cubic triangles takes the longest, which is unsurprising given the large number of optimziation variables. Lower-degree problems like MBB and MSS are already fast enough to be used on large meshes out of the box. 
    *For SI, 100\% exact recovery is taken out of cases where the patch did self-intersect. Exact recovery is not expected if the patch does not self-intersect.}
    
    \begin{tabulary}{2.0\columnwidth}{@{}LCCCCCCCCCCCCCC
    @{}}\toprule
    & \multicolumn{2}{c}{$\dcorr$} & \multicolumn{2}{c}{\% Correct} & \multicolumn{2}{c}{\% Exact} & \multicolumn{2}{c}{\% Intersects} & \multicolumn{2}{c}{$\#$ Vars} & \multicolumn{2}{c}{\textsc{yalmip} Time (s)} & \multicolumn{2}{c}{\textsc{mosek} Time (s)}\\%
    \cmidrule{2-3} \cmidrule{4-5} \cmidrule{6-7} \cmidrule{8-9} \cmidrule{10-11} \cmidrule{12-13} \cmidrule{14-15}
        Triangle $d_{\bfx}$ & {$2$} & {$3$} & {$2$} & {$3$} & {$2$} & {$3$} & {$2$} & {$3$} & {$2$} & {$3$} & {$2$} & {$3$} & {$2$} & {$3$}\\ \midrule 
        CP  & 3 & 5 & 100  & 100 & 100 & 100 & - & - & 31 & 64 &   .56$\pm$.04 & .52$\pm$.04 & .14$\pm$.02 & .10$\pm$.08 \\ 
        MBB & 2 & 4 & 100  & 100 & 100 & 100  & - & - & 19 & 46 & .48$\pm$.02 & .49$\pm$.02 & .002$\pm$.0003 & .004$\pm$.0007   \\ 
        PD  & 4 & 6 & 100  & 100 & 100 & 100   & - & -  & 491 & 1471 & .49$\pm$.02 & .5$\pm$.02  & .05$\pm$.009 & 1.24$\pm$.19  \\  
        SSI & 5 & 6 & 100  & 100 & 27 & 0   & 60 & 85 & 757 & 1261 & .6$\pm$.02 & .66$\pm$.08  & .78$\pm$.12 & 6.19$\pm$1.4 \\ 
        SI  & 4 & 4 & 100  & 100 & 100* & 0   & 51 & 84  & 491 & 491 & .38$\pm$.02 & .5$\pm$.03  & .29$\pm$.06 & .39$\pm$.09\\ 
        CCD & 5 & 6 & 100  & 100 & 1 & 0    & 82 & 90 & 2017 & 3697 & .69$\pm$.04 & 1.31$\pm$.09   & 25.9$\pm$4.4 & 699$\pm$49\\ 
        MSS & 2 & 4 & 100  & 100 & - & - & - & - & 36 & 63 & .41$\pm$.03 & .38$\pm$.008    & .006$\pm$.002 & .012$\pm$.002\\ 
        MEE & 4 & 6 & 100  & 100 & - & -      & - & - & 68 & 107 & .42$\pm$.04 & .37$\pm$.03    & .13$\pm$.02 & .13$\pm$.01   \\ 
        \bottomrule
    \end{tabulary}
    
    \label{tbl:combined}
\end{table*}
If the moment vector $\bm{\mu}$ solving \eqref{eq:lasserre} corresponds to a valid probability measure, then we know the relaxation was tight and we have achieved the global optimum of the original measure relaxation \eqref{eq:measureopt}, and in turn of the polynomial problem \eqref{eq:templateproblem}. Verifying this when the optimal measure is a delta measure is particularly simple---this will be the case if and only if the semidefinite matrix corresponding to the first constraint in \eqref{eq:lasserre} has rank one. This phenomenon is known as \emph{exact recovery}, and it provides an \emph{optimality certificate} verifying that $d$ was chosen sufficiently high for the relaxed problem to solve the original problem \eqref{eq:templateproblem}. Finally, when we have exact recovery, $\bfu^*$ can be recovered as the mean (i.e., vector of first moments) of $\delta_{\bfu^*}$: 
\begin{equation}
\bfu^* = (\mu_{\bfu_1}, ..., \mu_{\bfu_k})
\label{eq:momentmean}
\end{equation}

\paragraph{Revisiting PD and SI}
With \eqref{eq:momentmean} we are equipped to extract $\bfu^*$ for problems with a unique global minimum. Problems like PD, however, have multiple global minima due to the exchange symmetry $\bfu_1 \leftrightarrow \bfu_2$, which leaves the objective function unchanged. Such a symmetry means that the computed optimum $\bm{\mu}$, even if it corresponds to a valid measure, will not in general be a delta measure, making exact recovery elusive. Moreover, the mean extraction \eqref{eq:momentmean} will yield the Euclidean mean of optimal values, which may not itself be optimal.

To address these issues, we break the exchange symmetry by adding the generic constraint
\begin{equation} g(\bfu_1,\bfu_2) = (\bfu_1 - \bfu_2)\cdot \vec{v} \geq 0, \end{equation}
where $\vec{v}$ is a randomly sampled unit vector. With this modification, generic uniqueness of the $\argmin$ is restored, allowing for exact recovery.

SI has the same symmetry and requires the same symmetry-breaking constraint. Furthermore, exact recovery is only possible for SI if the patch has a self-intersection. Otherwise, any pair of points $\bfu_1=\bfu_2\in\D$ are globally optimal with an optimal value of 0. Despite the possible lack of exact recovery, we address in \Cref{sec:momentrelaxation:prac} how to solve problems like SI reliably in practice.

\begin{figure*}
\centering
\include{generateCodeFigureBody}
\vspace{-35pt}
\caption{Conversion from \eqref{eq:templateproblemsos} to functioning \textsc{yalmip} code for CP. The \textsc{yalmip} code matches the mathematical formulation almost line to line. After specifying the the primal SOS problem, the dual moment vector and $\bfu^*$ are easy to extract. To solve other problems on different patches, only \texttt{gi, trimapX} and \texttt{f} need to be changed.}
\label{fig:yalmipcode}
\end{figure*}

\subsection{Moment Relaxation in Practice}
\label{sec:momentrelaxation:prac}
A general algorithm to achieve exact recovery might be to incrementally increase the SOS degree $d$ from 1 until exact recovery is achieved. While this strategy may succeed eventually, \citet{marschner2020hexahedral} observe that moment relaxations of a specific problem class tend to achieve exact recovery at the same degree, independent of the specific numbers involved. In particular, they observe that for a trilinear hexahedron, the hexahedron validity problem consistently achieves exact recovery with $d=4$. Given a problem type, one might determine $d$ empirically by generating many randomized instances of the same problem and increasing $d$ until these  generic instances can be solved to exact recovery. Then the empirically-determined degree can be applied to untested instances of the same problem.

Curiously, we observe that it is frequently possible to obtain $f^*$ and $\bfu^*\in\D$ even without exact recovery. When $d$ is smaller than required for exact recovery, the moment vector $\mu$ does not encode a delta measure. Nevertheless, we can take its mean following \eqref{eq:momentmean} and treat that as a proxy for $\bfu^*$. While this leaves the realm of formal SOS theory, we find in practice that this strategy is extraordinarily robust. Consider the SSI problem on quadratic triangles in \Cref{tbl:combined}. Choosing degree $d = 5$ provides us with exact recovery for only 27\% of the instances of this problem. However, we can correctly detect intersection 100\% of the time, and our extracted $\bfu^*$ gives us the correct intersection points 100\% of the time. The same phenomenon is observed for SI and CCD---exact recovery is not required to obtain the correct solution.

Thus we distinguish the degree $\dexact$ at which exact recovery is achieved from the degree $\dcorr$ at which a correct solution is achieved. In problems like SSI where numerical verification of intersection is easy given the candidate points of intersection, having an accurate $\bfu^*$ is just as effective a certificate of intersection as exact recovery.
Since the cost of solving an SOS program is primarily dictated by $d$, it is extremely fortunate that for practical purposes one does not require exact recovery.

We demonstrate in \Cref{fig:degrading} the percentage of correctly solved SSI problems out of 200 for each of $d\in{1,2,3,4,5}$ using randomly sampled quadratic triangles. Our sampling strategy is described in \Cref{sec:results_overview}. 128 of these problems have intersections. Out of problems with intersections, we plot the percentage of them that exhibit exact recovery for each degree. This percentage never reaches 100\% and our key observation is that it does not need to! We do not require exact recovery to obtain the correct solution. We plot the percentage of problems where the SOS formulation misidentifies whether an intersection exists. At degree only $d=2$, the percentage misidentified is already zero. Thus we are already able to determine whether a pair of quadratic triangles intersect with $100\%$ accuracy. Finally, for problems that have intersections, using \eqref{eq:momentmean} we can extract the points on each surface that intersect. When the SSI problem is correctly solved, their distance must be 0. We plot the maximum distance between the extracted points over all intersecting examples at each degree.
At $d=5$, the points of intersection are identified with high precision. While these problem instances do not exhaustively cover SSI, they provide very strong empirical evidence that exact recovery is not required.

\section{Results and Applications}
\label{sec:results}

\subsection{Implementation}
Modern problem modeling languages such as \textsc{yalmip} \cite{Lofberg2004} allow one to specify an SOS program in the concise form \eqref{eq:templateproblemsos}, automatically converting it to a primal-dual SDP formulation from which one can obtain the moment vector $\mu$ of \eqref{eq:lasserre}. \Cref{fig:yalmipcode} demonstrates how easy it is to specify and solve the primal and dual problems. The details of converting SOS or moment constraints into an SDP can be entirely left to the modeling language. 

All batch experiments were run on an Intel i7-8700K CPU @ 3.70GHz with 16 GB of RAM. Problems were all written in \textsc{matlab} 2021a using \textsc{yalmip} version 20200116 to formulate the problems and \textsc{mosek} 9.2 to solve. Default modeling and solver parameters were maintained. \Cref{tbl:combined} lists the median runtimes and standard deviations for 50 random instances of each problem for quadratic and cubic triangles. Time spent in \textsc{yalmip} converting the problem into a form that \textsc{mosek} can handle does not vary significantly across problems or patch types. Time spent actually solving the problem in \textsc{mosek} varies much more. Clear trends are that solver time increases for each problem with SOS degree, which increases with patch degree.
While CCD seems to be a significantly challenging problem, MBB is fairly easy. For most problems, the table shows that the majority of the time is spent in \textsc{yalmip}. This time can be cut out entirely by formulating the problem directly as an SDP problem, or by taking advantage of \textsc{yalmip}'s ability to precompile problems. We expect the runtime can be further improved by engineering tailor made optimization strategies for these problems.


\subsection{SOS Problems Overview}
\label{sec:results_overview}
A key parameter of our SOS formulations is the degree $d$. Since it controls the size of the SDP in \eqref{eq:SOSSDP}, we naturally want to choose a small $d$. However, if $d$ is too small, the SDP may not successfully solve the unrelaxed problem. In \Cref{tbl:combined}, we show the minimal degree $\dcorr$ required to solve each problem for 100\% of our test cases on various triangular patches.
As mentioned in \Cref{sec:momentrelaxation:prac}, this is not necessarily the same degree $\dexact$ needed to achieve exact recovery.

For each problem, 1000 test instances were generated, except for CCD on cubic triangles, for which we used 300 instances. For CP, MBB, PD, SI, SSI, MSS, and MEE, we randomly sampled quadratic and cubic triangles by picking all dimensions of all control points independently according to a normal distribution with variance 1. For CCD, pairs of triangles are sampled the same as before, but then shifted by  $(0,0,\frac{\sqrt{2}}{2})$ and $(0,0,-\frac{\sqrt{2}}{2})$ respectively. Velocity vectors are sampled the same way, but shifted by $(0,0,-\frac{\sqrt{2}}{2})$ and $(0,0,\frac{\sqrt{2}}{2})$ respectively. This is done to increase the number of problem instances that successfully collide without starting out in an intersecting configuration. With this sampling strategy, 82\% (90\%) of the sampled quadratic (cubic) triangles collide in spacetime while 22.5\% (28\%) of configurations start out intersecting. We verify correctness of the SOS solutions by comparing to the same problem solved on a linearization of the polynomial patch. Patches are uniformly subdivided so that each edge is split into 10 segments. We then check that solutions for the SOS solution and the linearization match up to a threshold of $10^{-2}$ in both the parametric and embedded spaces. In a few cases where they do not match, we increase density of the linearization until they do.

\Cref{tbl:combined} demonstrates how challenging it can be to guess $\dcorr$ in advance. The SOS degrees we end up with are not clearly correlated with the degrees of the objective or constraints. The clearest trend is that $\dcorr$ increases with the degree of the shape function $d_{\bfx}$. There does not seem to be a monotonic relationship between $\dcorr$ and the degree of the objective function. For example, CP has an objective degree of $2 d_{\bfx}$ while needing fairly low $\dcorr$. On the other hand, CCD has an objective degree of just 1 but needs a much higher $d$. Even if we consider the constraint degrees as well, the degree of the CCD constraint is only $d_{\bfx}+1$ which is still less than CP's objective degree $2 d_{\bfx}$. 

\subsection{Closest Point}
\begin{wrapfigure}{r}{0.40\columnwidth}
    \centering
    \includegraphics[width=0.40\columnwidth]{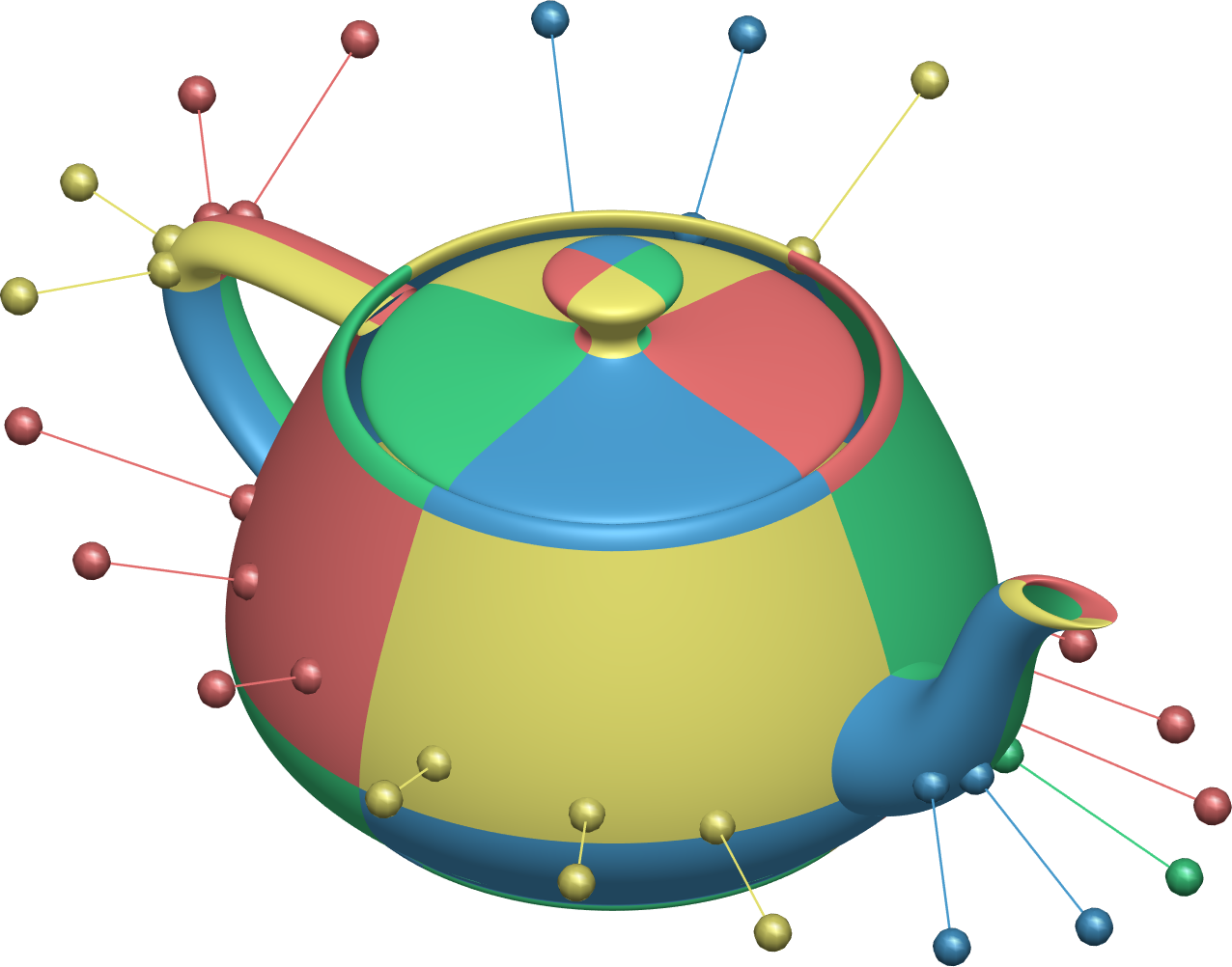}
\end{wrapfigure}
As described in \ref{sec:singlepatch}, our method of formulation is applicable to the problem of finding the closest point on a polynomial patch to a target point.

\begin{figure}
    \centering
    \includegraphics[width=1\columnwidth]{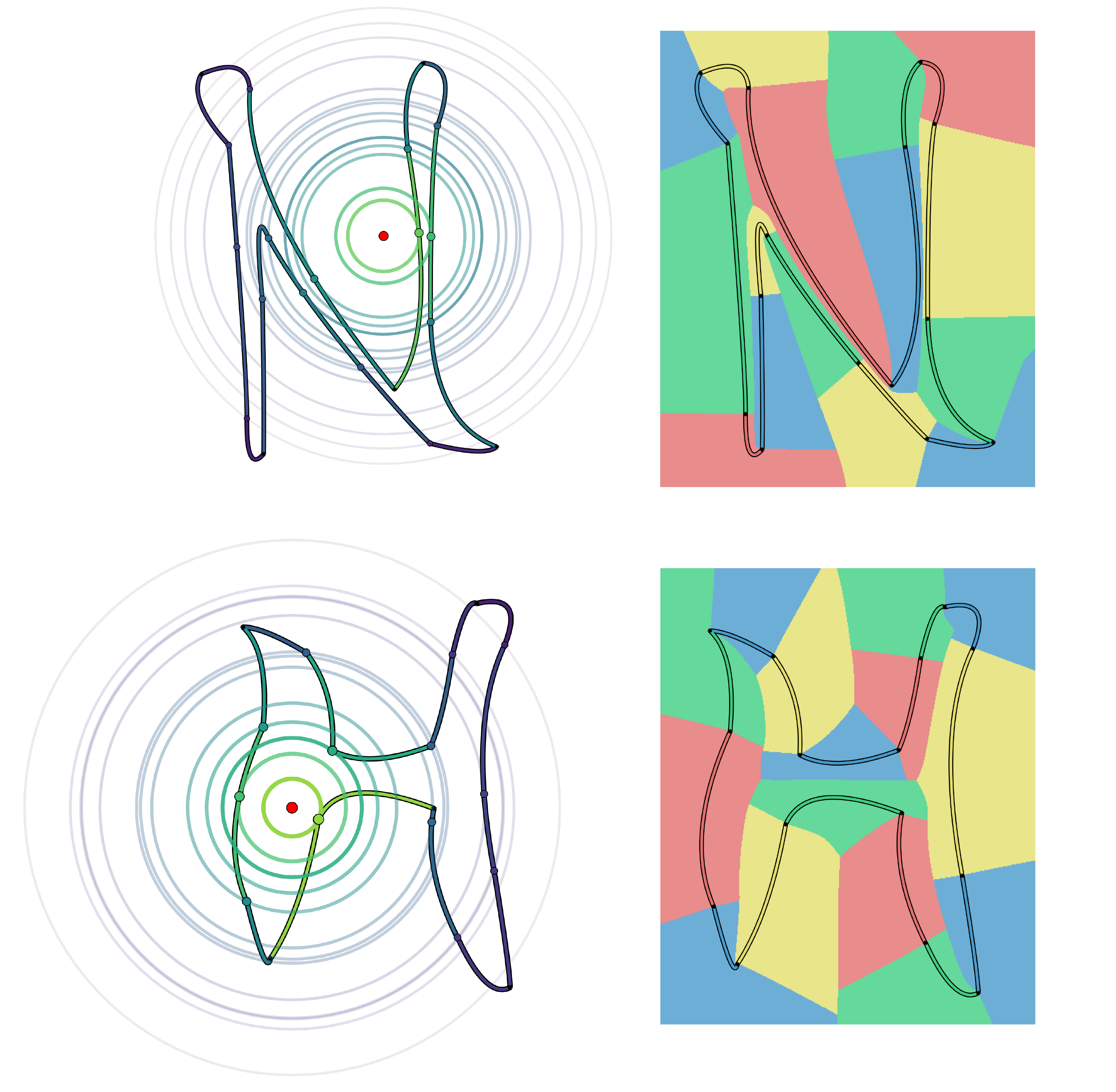}
    \vspace{-25pt}
    \caption{Curve segment Voronoi diagrams for the glyphs H and N from \cite{smirnov2020deep}, each composed of 15 quadratic B\'ezier curves. On the left, we show the calculation of the closest point on each of the curve segments to the red point, with the distance to each curve represented by a circle. On the right, we show the Voronoi diagram, computed by finding the curve with the minimum distance to each point in a 500 by 500 grid around each letter.}
    \label{fig:letter_voronoi}
\end{figure}

Applying this method for all patches in a model lets us find the closest point  on the model to a target. In the inset figure, we show this procedure applied to a collection of points surrounding the bicubic Bézier tensor patch teapot model, where $d_{\bfx}=6$ and we chose $d=5$.
We can solve an analogous problem in one lower ambient dimension, finding the closest point to a Bézier curve in $\mathbb{R}^2$.
This formulation allows for the creation of Voronoi diagrams of 2D objects formed from B\'ezier curves, as shown in \Cref{fig:letter_voronoi}. In this figure, the 2D objects are letters consisting of quadratic Bézier curves, where $d_\bfx=2$ and $d=3$. We color points in the 2D domain by which curve segment they are closest to, forming curve-segment Voronoi cells. 

\setlength{\intextsep}{2pt}
\setlength{\columnsep}{5pt}
\begin{wrapfigure}[10]{l}{0.3\columnwidth}
    \centering
    \includegraphics[width=0.3\columnwidth]{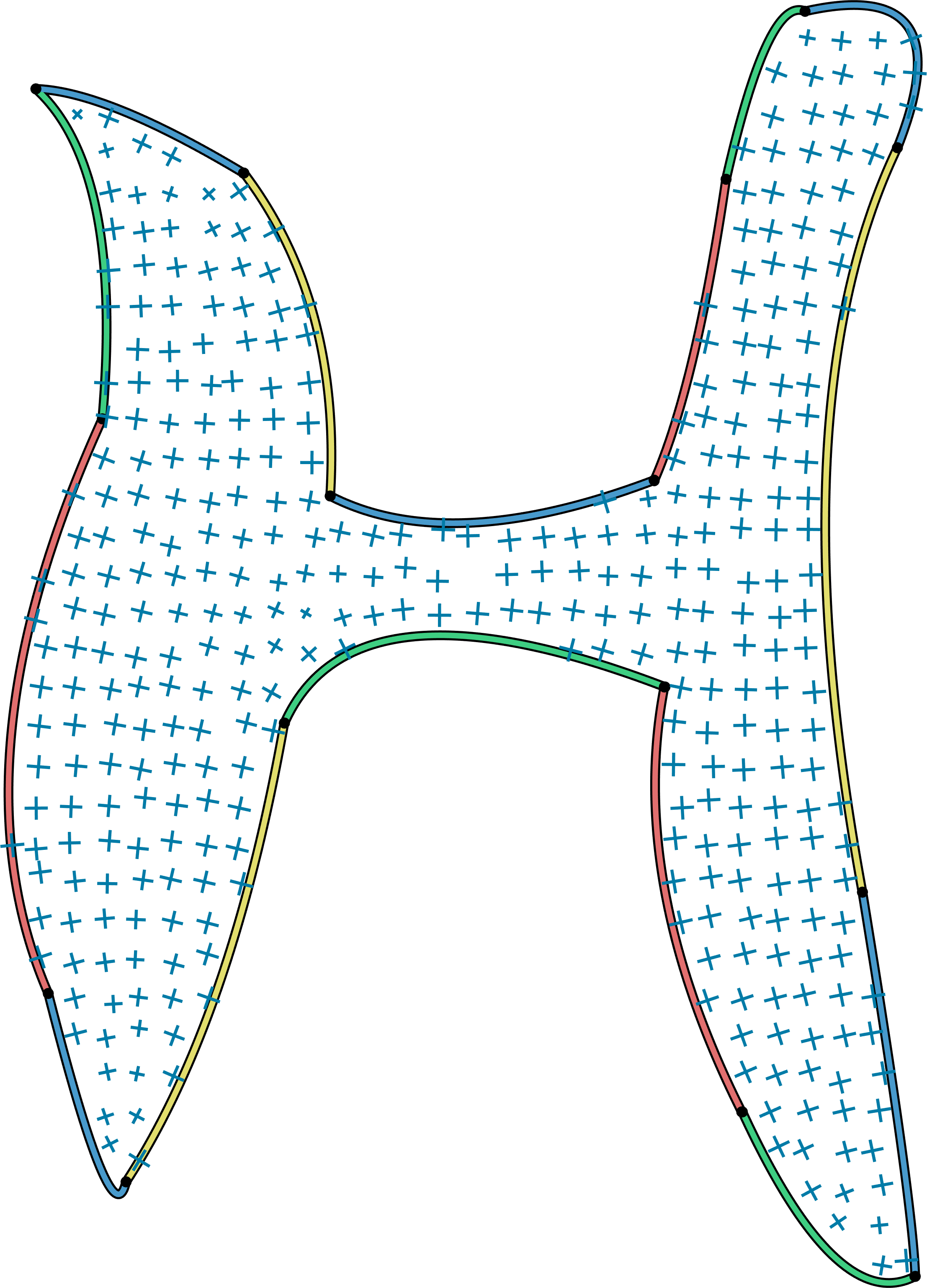}
\end{wrapfigure}
Closest point queries allow us to build more complex operations on domains with curved boundaries. For example, the Monte Carlo walk-on-spheres procedure introduced to geometry processing in \cite{Sawhney:2020:MCG}  builds a stochastic solver for linear PDE out of closest point queries.
Using the SOS formulation of CP as a building block, such a procedure can solve a PDE on a domain defined by polynomial bounding curves with exact boundary conformation, without linear remeshing. In the inset figure, we demonstrate walk-on-spheres applied to computing a boundary aligned cross-field. Each frame is computed by averaging complex fourth powers sampled from $100$ random walks, each stopping when it arrives within $10^{-3}$ of the boundary. No mesh of the domain interior is required, and in principle, samples can be adapted to resolve singularities precisely.

\begin{figure*}
    \centering
    \newcommand{\imgwidth}{0.4\columnwidth}
    \begin{tabulary}{\textwidth}{CCCC}
    \includegraphics[width=0.8\columnwidth]{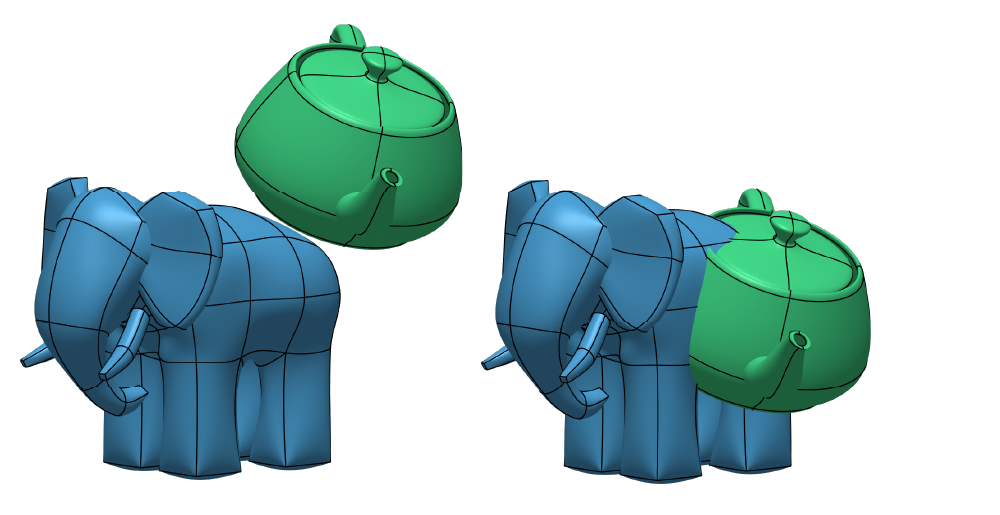} &
    \includegraphics[width=\imgwidth]{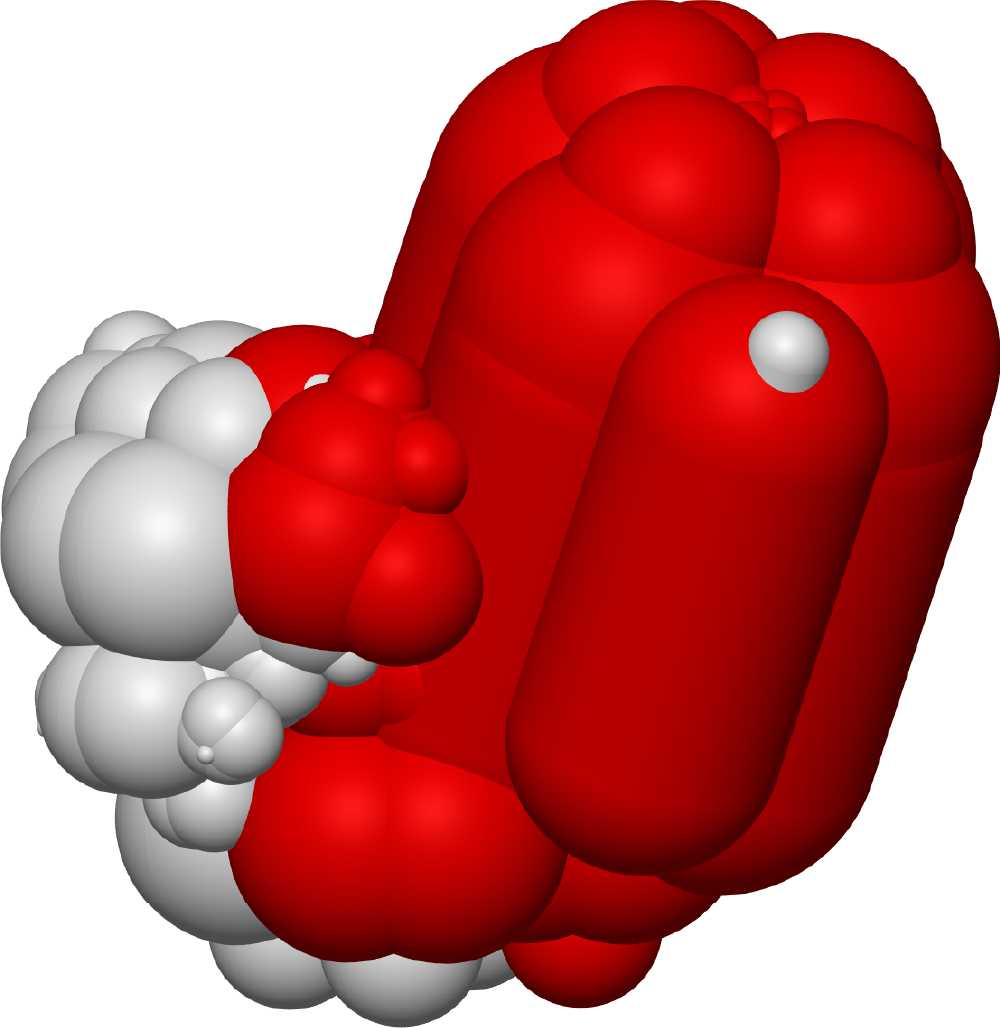} &
    \includegraphics[width=\imgwidth]{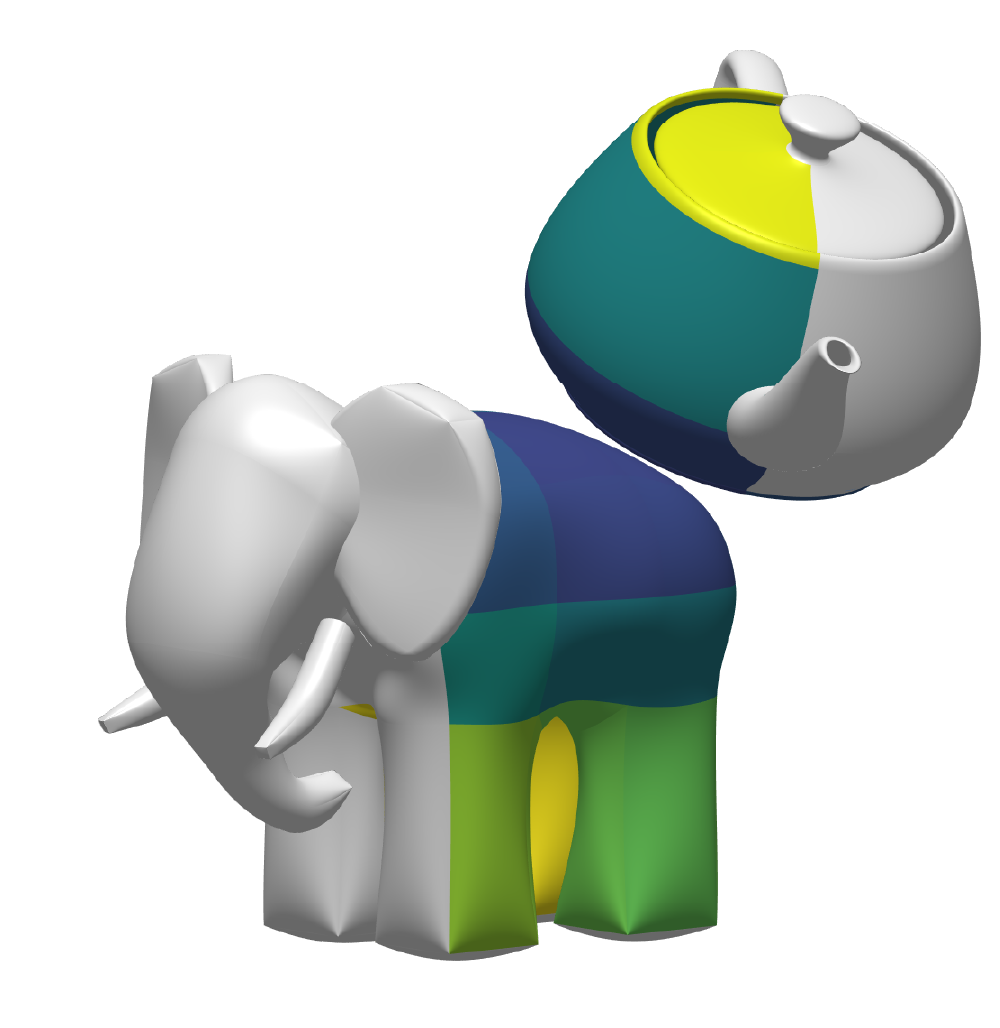} &
    \includegraphics[width=\imgwidth]{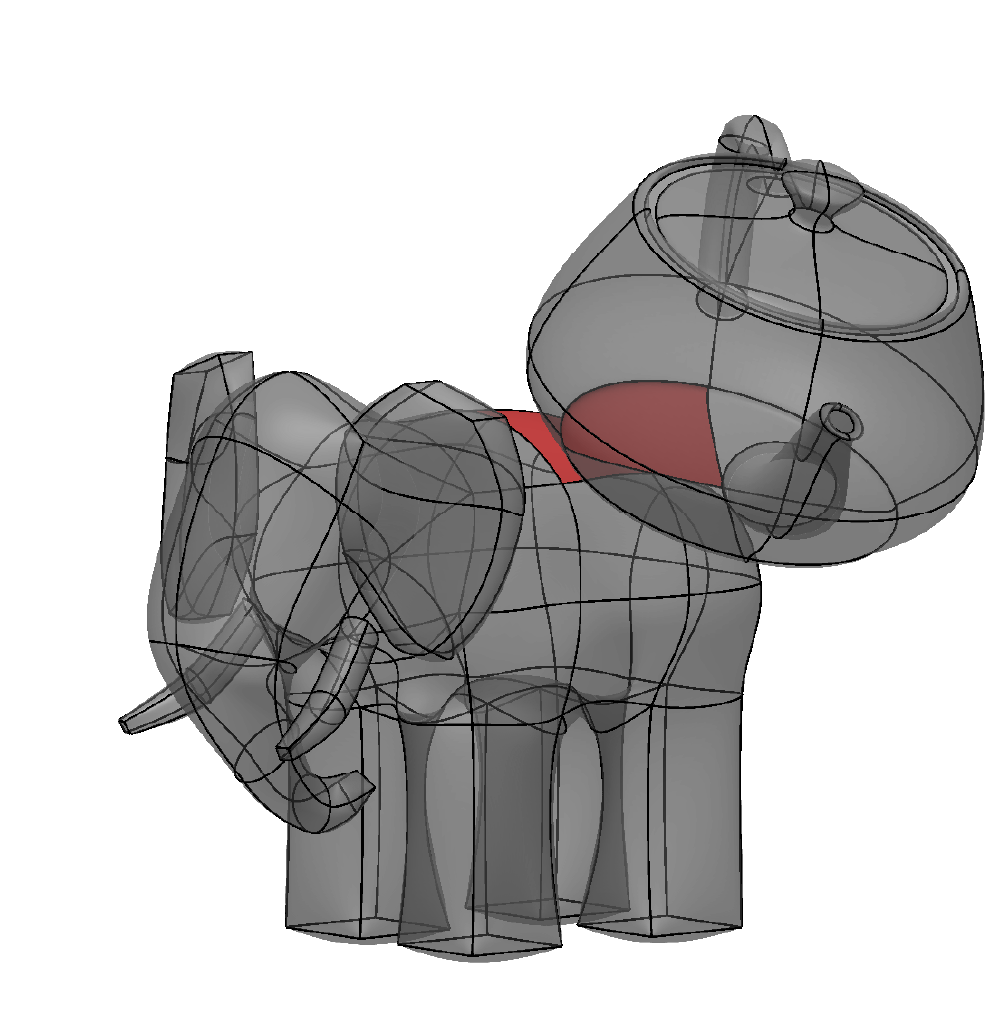}\\
    (a) Problem Setup & (b) Bounding Sphere CCD & (c) Exact CCD & (d) First Intersection
    \end{tabulary}
    \vspace{-5pt}
    \caption{We show the application of our CCD solution to the problem shown in (a), where the teapot moves rigidly into the elephant. The teapot and elephant models are comprised of 32 and 128 bicubic Bézier tensor patches, respectively. To accelerate the calculation of the collisions, we first use our MSS method to pre-compute a set of bounding spheres for each model, and use these to determine which pairs of patches may intersect. In (b) we color in red the bounding spheres of the elephant and bounding capsules of the teapot for which at least one intersection was found. We then solve the SOS patch CCD problem on the remaining candidate intersections.
    The results of this are shown in (c), where each patch is colored based on the time at which it first collides (with darker colors representing earlier times). We can also determine the exact moment in time when the two models first intersect, shown in (d).}
    \label{fig:ccd}
\end{figure*}

\subsection{Intersections}

It is worth revisiting the intersection problems (SSI, SI, CCD) in \Cref{tbl:combined}. 
First, for SI on quadratic triangles, $\dcorr=\dexact$. The same property does not extend to cubic triangles or SSI on quadratic triangles since the percentage that achieve exact recovery is different from the percentage that have intersection.

An interesting feature of SI is that the degree $\dcorr$ is the same for both quadratic and cubic triangles. This is surprising given that for other problems the cubic triangles require higher $\dcorr$. We note here that CCD was almost the same in that 93\% of cubic triangle CCD instances succeeded using $d=5$ with runtime comparable to CCD for quadratic triangles. Only for the remaining percentage was $d=6$ necessary. Due to the large time cost of CCD on cubic triangles with $d=6$, one can imagine an optimization where $d=5$ is used first to prune cases where a collision happens and is certifiably found. Only in the absence of this collision would one resort to $d=6$.

\begin{figure}
    \centering
    \includegraphics[width=1\columnwidth]{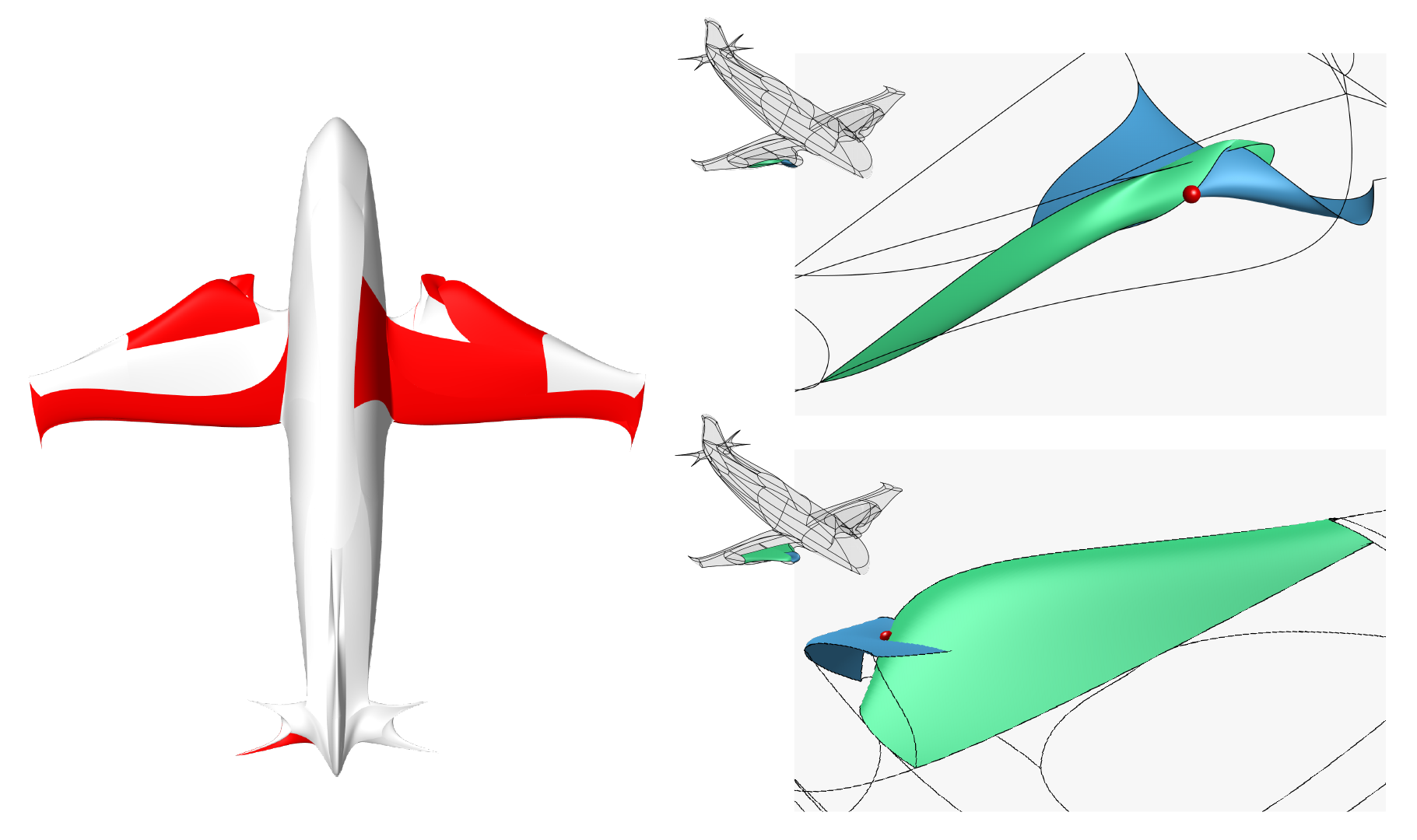}
    Surface-Surface Intersections
    \vspace{10pt}

    \includegraphics[width=1\columnwidth]{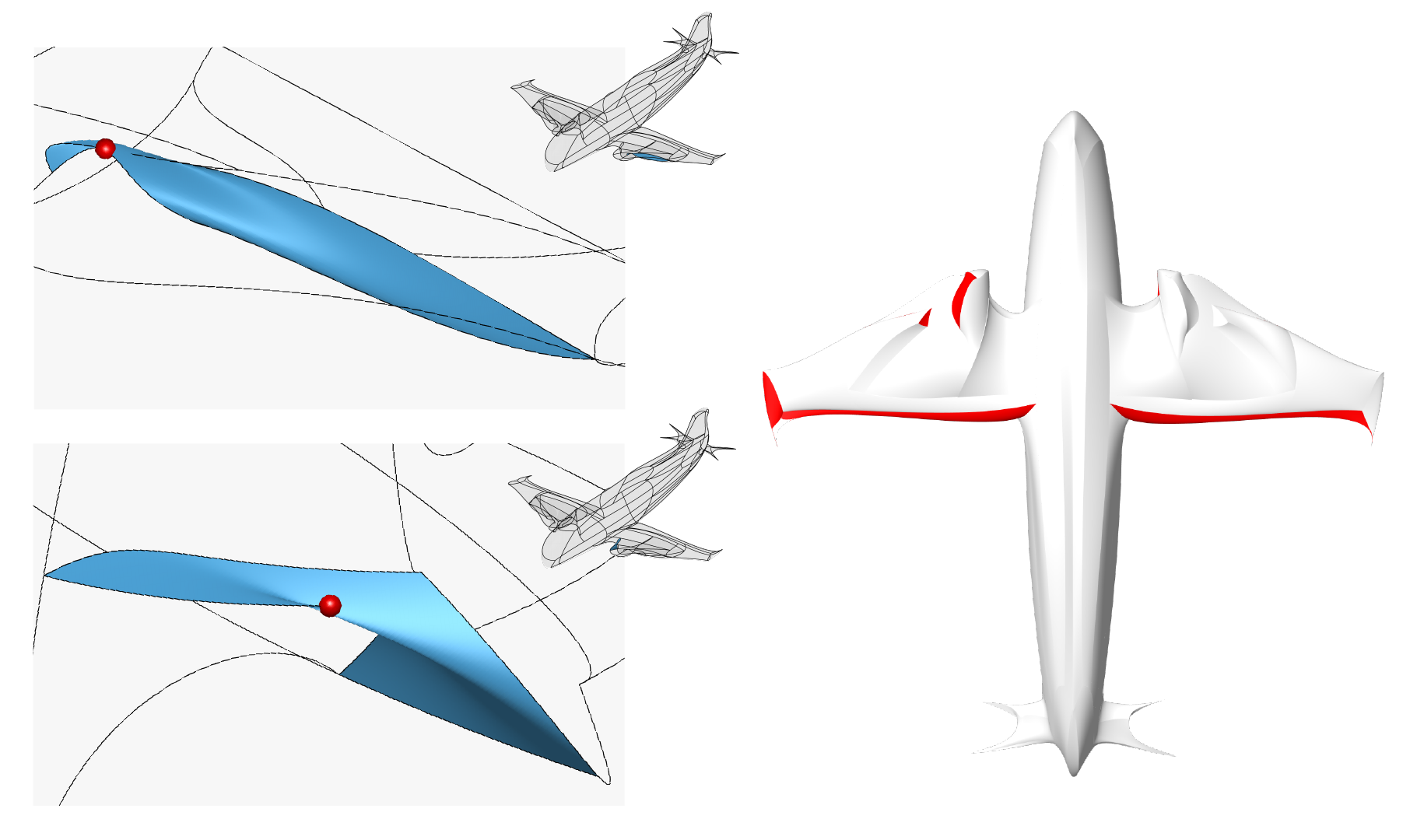}
    Self-Intersections
    \vspace{-5pt}
    \caption{Our method for finding self-intersections and surface-surface intersections applied to an invalid airplane mesh, obtained from the pipeline of \cite{smirnov2020learning}. In the top row, we show on the left all the patches that intersect with another patch in red. On the right, two of those surface-surface intersections are shown, with a red point marking the intersection point calculated by our method. In the bottom right, all the patches that include self-intersections are shown in red. Two of these self-intersections are shown on the left, with the intersection points plotted in red.}
    \label{fig:airplane_intersection}
\end{figure}

In \Cref{fig:airplane_intersection} we demonstrate application of SI and SSI tests to airplane meshes from \citet{smirnov2020learning}. These models use cubic Coons patches for which $d_{\bfx}=4$. The SI and SSI problems are solved with $d=4$. We are able to find intersections for both SI and SSI and provide their exact locations. This enables users in a CAD pipeline to discover and manually fix problems in a design.

In \Cref{fig:ccd} we demonstrate application of the SOS relaxed CCD problem to detect collision of a rigid bicubic Bézier tensor patch teapot and elephant. 
Velocities were chosen to guarantee collision within 1 unit of time. This problem was solved with $d=4$. Using CCD we are able to find the earliest instance of collision, the patches that collide, and the location of their collision. With the location of the collision we can get exact surface normals at the collision point, which can then be passed onto later steps in a simulation pipeline such as collision response.

\subsection{Bounding Volumes}
\begin{figure}
    \centering
    \includegraphics[width=1\columnwidth]{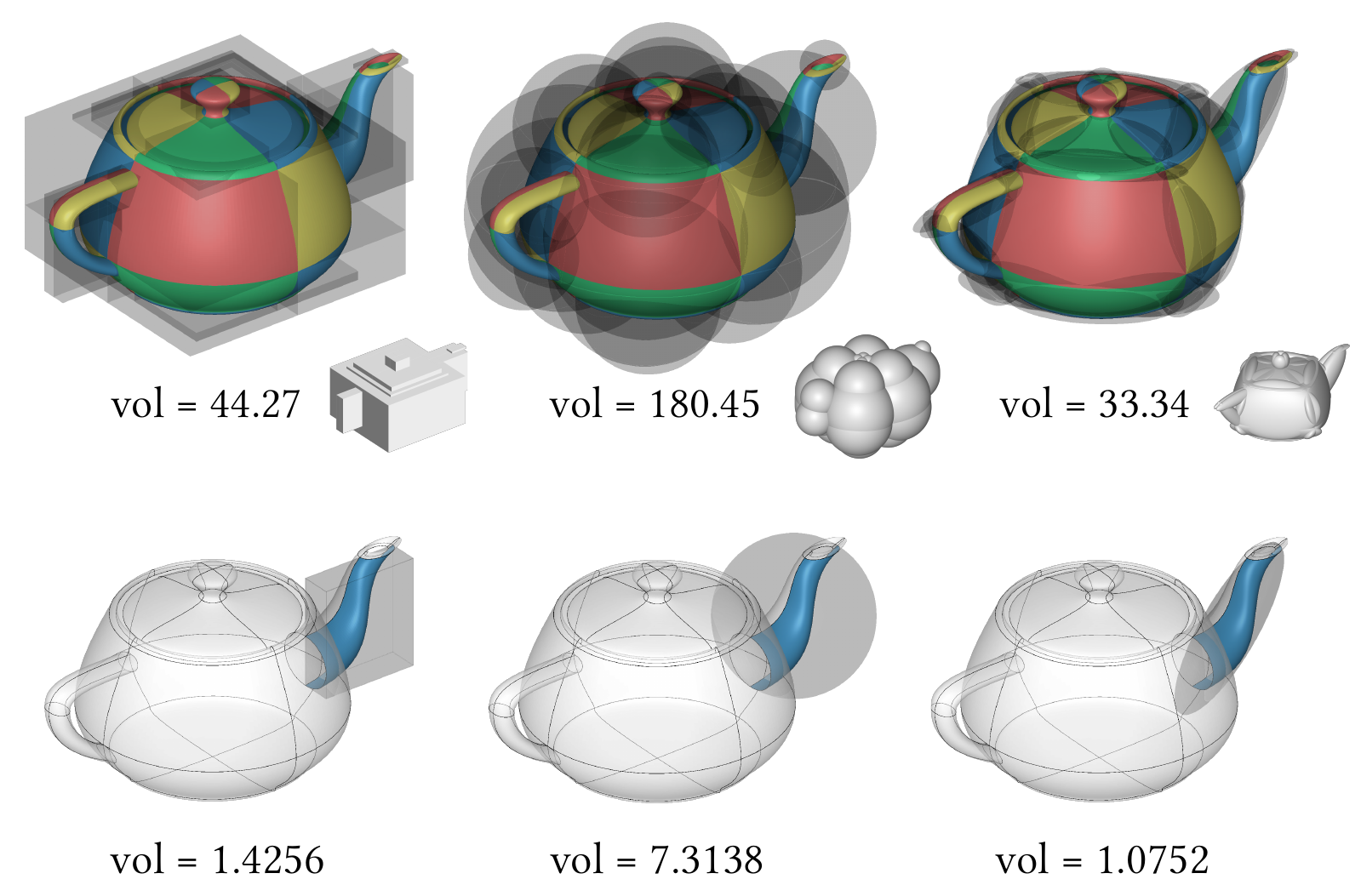}
    \vspace{-10pt}
    \caption{In the top row, we show a comparison of the MBB, MSS, and MEE for the bicubic Bézier tensor patch teapot. The total volume of bounding elements is computed for each type; ellipsoids provide the tightest bounding volume. The bottom row shows a comparison of the three different bounding types for an individual patch.}
    \label{fig:bdd_teapots}
\end{figure}

Bounding volumes are useful to speed up various intersection-type problems. They can be used in ray tracing to quickly detect if a ray will not intersect an object. They can also be used to quickly detect if rigid objects will not collide. The tighter a bounding volume is, the more non-collisions/intersections can be quickly pruned. On the other hand, employing a more complex bounding volume shape can result in increased computation and updating costs in cases when bounding volumes cannot be precomputed. As such, there is a tradeoff between the computation saved by having a tighter bounding volume and the cost of maintaining the bounding volume itself. SOS optimization is flexible enough to target multiple points along this tradeoff e.g. MBB, MSS, and MEE. In \Cref{fig:bdd_teapots} we compute the MBB, MSS, and MEE evaluated on bicubic Bézier tensor patches of the teapot mesh. As expected the total volume of MEEs is less than that of MSSs or that of MBBs.
The MBB problem fits into the framework of \Cref{sec:singlepatch}, and so we can generically expect exact recovery for high enough $d$. The MSS and MEE problems are different in that the MSS (MEE) generically intersects the polynomial patch it bounds at multiple points. Thus we do not ever expect exact recovery. Nevertheless, we can obtain the minimal volume and the parameters of the MSS (MEE) achieving that volume from \eqref{eq:cmee_quadmod}.

In \Cref{fig:ccd}, we demonstrate the MSS on rigid bicubic Bézier tensor patches as a way to shorten computation time of CCD problems between meshes. Instead of running CCD between all pairs of patches, we can drastically cut down computation time by precomputing the MSS for each patch. It is straightforward to check if two moving spheres will collide or not, allowing us to reduce the number of CCD computations required.
\begin{figure}
    \centering
    \includegraphics[width=0.4\columnwidth]{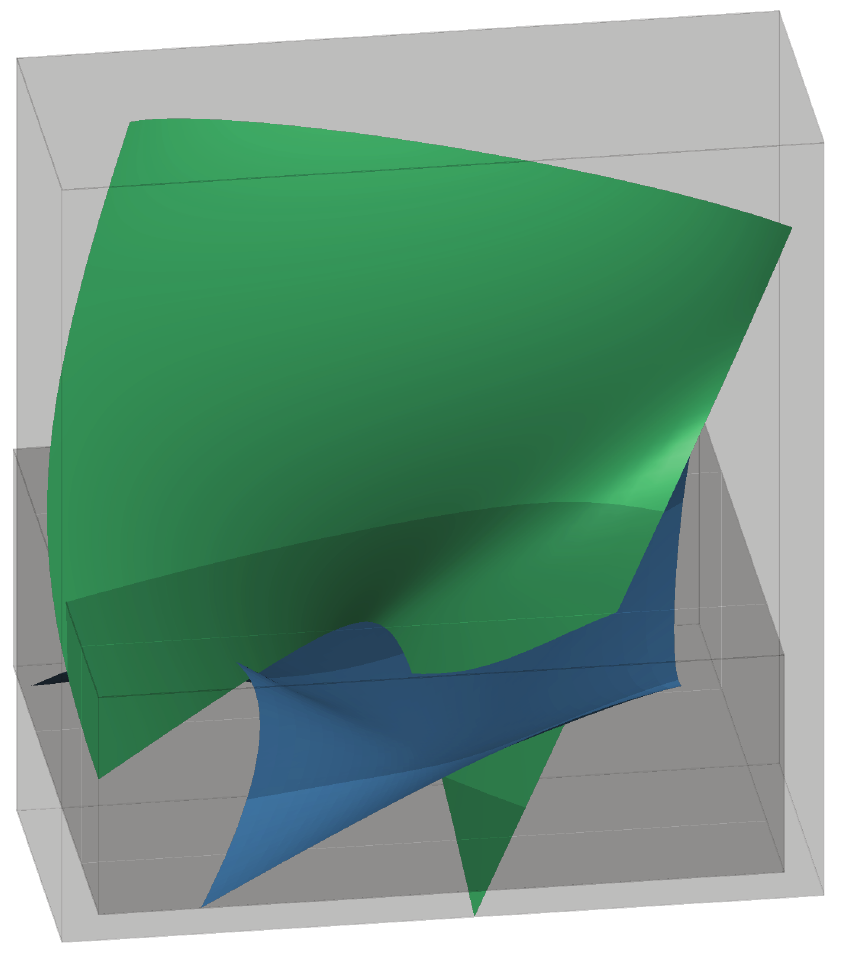} \hspace{0.1\columnwidth} \includegraphics[width=0.372125\columnwidth]{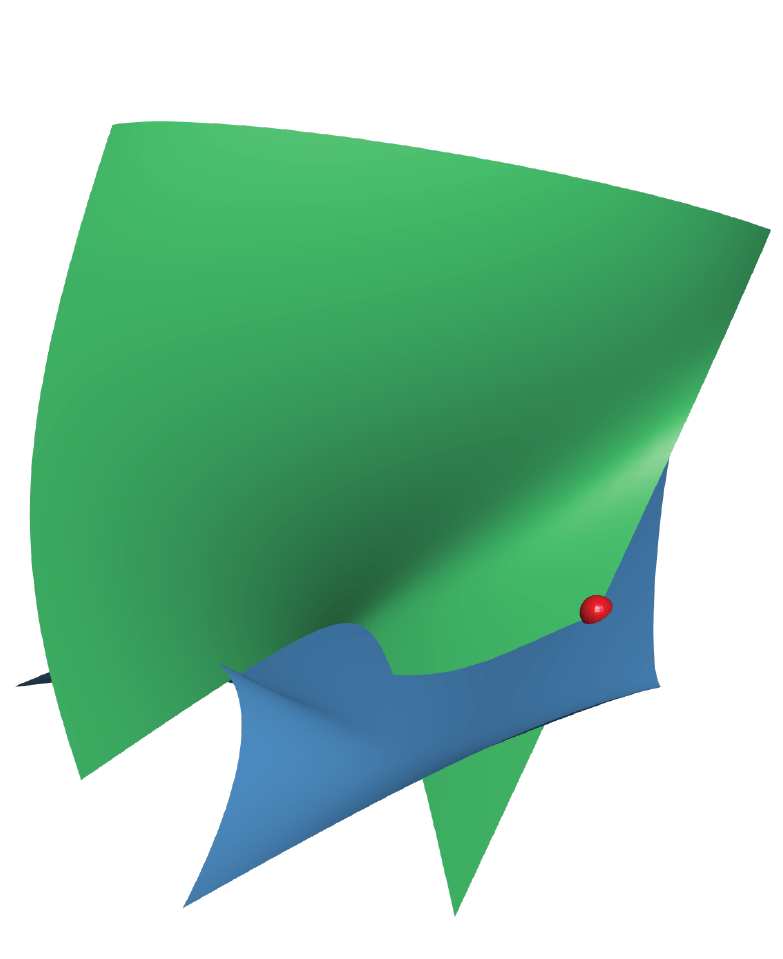}
    \vspace{-5pt}
    \caption{The MBB and SSI problems solved on generic NURBS patches. On the left, two bounding boxes are drawn around the blue and green patches. On the right, the red point indicates where these patches intersect.}
    \label{fig:actuallynurbs}
\end{figure}

\subsection{Rational Surfaces}
\begin{figure}
    \centering
     \includegraphics[width=1\columnwidth]{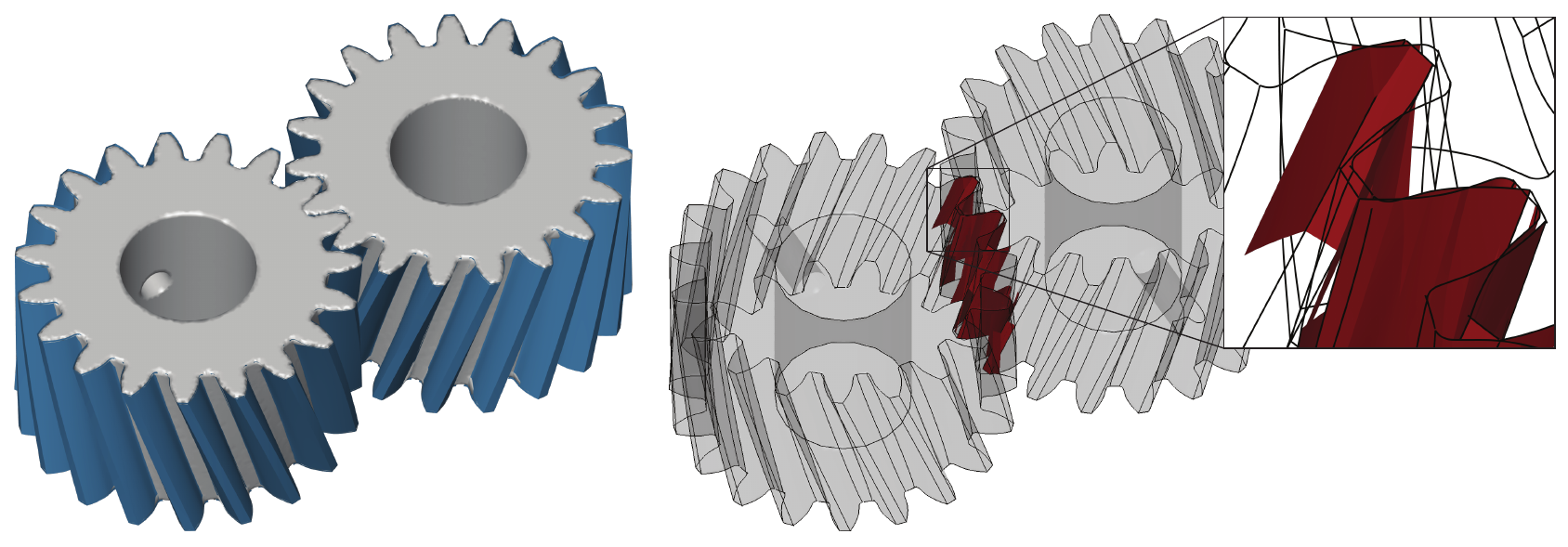}
    \caption{This helical gear, from the ABC dataset \cite{koch2019abc}, is modeled with NURBS surfaces defining its teeth, shown in blue in the left figure. To determine whether the two gears intersect, we can apply our SSI formulation to the NURBS surfaces in the two gears. To speed up the runtime, we first compute the bounding boxes of each patch with our MBB method and use these to identify patches which may intersect. The patches in the two gears that intersect are shown in red on the right. }
    \label{fig:nurbsgear}
\end{figure}

We demonstrate application of SSI to NURBS in \Cref{fig:nurbsgear} on gears obtained from \cite{koch2019abc}. Intersecting teeth are easily detected and highlighted. We also demonstrate CCD in \Cref{fig:teaser} on the castle from \cite{Trusty_Chen_Levin_SEM_2021} to find which patches of the rocket will first collide with patches of the castle. These collisions could then be fed back into their elasticity simulation to render colliding elastic objects without need for linearization or volumetric re-meshing.

While handling of rational surfaces allows us to process generic NURBS surfaces, we find that many of the models in \cite{koch2019abc} and \cite{Trusty_Chen_Levin_SEM_2021} have mainly patches with constant denominators $b=1$, making them simply B-Splines. The primary exceptions in these datasets are patches that comprise spherical or cylindrical features. In \Cref{fig:actuallynurbs}, we show MBB and SSI on generic rational surfaces without constant denominator computed using SOS programming.

\section{Discussion and Open Problems}\label{sec:discussion}
This paper presents the use of SOS programming tools to solve a variety of problems in geometry processing. 
The miraculous success of SOS for geometry processing---beyond the bounds specified by the SOS theory---leaves many open questions for theory to answer. We document these questions below to inspire additional research developing SOS theory relevant to our new problem domain.

\vspace{-5pt}
\subsection{Advancing SOS Methodology}
\paragraph{Exact Recovery} A surprising aspect of our results is the robustness of the obtained solutions even when they do not come with exact recovery. This gap between theory and practice suggests formulating a modified notion of exact recovery where atomic measures are not required. It also suggests that there could be alternative formulations for the SSI, SI and CCD problems where exact recovery is found at the same degree that the correct solution is obtained.

For example, we relied heavily on Putinar's Positivstellensatz \Cref{thm:putinar} for our SOS formulations in this paper. While this form of the Positivstellensatz provides a concise form for certificates for compact domains, lower-degree certificates can sometimes be found by employing the full Positivstellensatz, which incorporates products of constraints \cite{blekherman2012semidefinite}.

\paragraph{Hardware}
Despite the flexibility of SOS programming, the runtime cost of solving SDPs can be substantial. 
Similar to how the availability of high-performance GPUs catalyzed deep learning, we expect that design of specialized SDP hardware would enable the use of SOS programming in many more contexts where runtime was previously too large. At the least, individual kernel problems do not interact and so can be parallelized.

\subsection{Exact Recovery in SI and CSI}

\paragraph{Self Intersection with Tolerance} One might also look for a version of SI that generically has a unique optimum when no intersection is present. Recall that for SI we chose the objective $f(\bfu) = -\|\bfu_1-\bfu_2\|_2^2$ to maximize distance in $\D$ when the SI problem only really cares about $\bfu_1\neq\bfu_2$. This suggests an alternative formulation with constraint $g(\bfu_1, \bfu_2) = \|\bfu_1-\bfu_2\|_2^2 - \epsilon \geq 0$, thus freeing the objective slot to be filled with something generic like $f(\bfu_1,\bfu_2) = (\bfu_1-\bfu_2)\cdot \vec{v}$ for a random unit vector $\vec{v}$. Now feasibility of the problem decides if there is or is not an intersection, while exact recovery can be expected in either case. The cost of this formulation is of course the addition of the tuning parameter $\epsilon$, and one might be left wondering if an undetectable self-intersection could have occurred in less than $\epsilon$ distance in $\D$.

\paragraph{Continuous Self-Intersection (CSI)} A secondary benefit of reformulating SI with tolerance $\epsilon$ is that it would allow us to solve the CSI problem. Given control points and velocities of a single polynomial patch, one looks for the earliest time that a self-intersection occurs. Now that the objective slot is vacated, it can be used to minimize $t$ similarly to CCD. The same drawback applies of course---$\epsilon$ must be chosen in advance.

\subsection{Degenerate Instances and Generalized Exact Recovery}
Our discussion of exact recovery has focused on the case where the optimal moments correspond to a delta measure, i.e., where the moment matrix has rank one. For some problems, optimal measures generically have more than one support point; this is the case for MSS and MEE, for which optimal measures are supported on points where the patch contacts the surface of the bounding sphere or ellipsoid. PD and SI would exhibit similar behavior if not for our symmetry-breaking modification. Other problems, such as CP, have \emph{degenerate instances}, i.e., problem instances for which there are multiple global optima to the polynomial optimization problem. For these instances, optimal solutions to the measure relaxation are mixtures of these global optima.

In such cases, one might still be interested in extracting the support points. While more complicated than in the rank-one case, it is still possible to extract an \emph{atomic measure}---a mixture of deltas---corresponding to a moment matrix when it satisfies the so-called \emph{flat extension} property. This property holds when increasing the degree of the moment relaxation does not increase the rank of the moment matrix. When flat extension holds, one can apply a procedure based on diagonalization of a set of commuting multiplication operators to extract the measure support points \cite[\S3.5.6]{blekherman2012semidefinite}. It would be interesting to test whether our MEE and MSS relaxations satisfy the flat extension property, which would serve as a certificate of generalized exact recovery and allow for extraction of the support points.

\subsection{Optimization with Quantifiers}
\label{sec:alternatequant}
Certain problems are difficult to manipulate into the form of the template problem due to the presence of \emph{quantifiers}. Quantifiers occur when the problem is more naturally formulated as a minimax or nested optimization problem. 
An example of this is MEE \eqref{eq:meefull}. While we managed to manipulate the MEE problem into a convex form, there are similar problems with quantifiers for which convex relaxations are still unknown. One example is the hexahedron repair problem presented in \cite{marschner2020hexahedral}, where they aim to minimally perturb a trilinear map such that the map becomes injective. Their alternating optimization solution, while effective in practice, comes with no global optimality guarantees.

\paragraph{Continuous Collision Response (CCR)}
Another example of an optimization with quantifiers is CCR:
given velocities $\hat{\bfv}_i^1$ and $\hat{\bfv}_i^2$ for two patches $\bfx_1$ and $\bfx_2$, find new velocities close to the input such that the new velocities do not result in collision in time $t_{max}$. Using the same notation as from CCD, we can write the CCR problem as the following.
\begin{equation}
\begin{aligned}
\label{eq:CCR}
v^* = \left\{
\begin{aligned}
& \underset{\bfv_i^1,\;\bfv_i^2\in\R^3}{\text{argmin}} & & \sum_{i=1}^{n_B} \|\bfv_i^1-\hat{\bfv}_i^1\|^2 + \|\bfv_i^2-\hat{\bfv}_i^2\|^2 \\
& \text{s.t.} & & 
\epsilon \leq 
\left\{\begin{aligned}
& \underset{\textbf{u}^1,\;\textbf{u}^2,\;t}{\text{min}} & & \|\bfx^1(\bfu^1,t) -  \bfx^2(\bfu^2,t)\|^2_2 \\
&\text{s.t.} & & t \in [0, t_{max}]\\
& & & \bfu^1 \in \D_1 \\
& & & \bfu^2 \in \D_2
\end{aligned}\right\}
\end{aligned}\right\}
\end{aligned}
\end{equation}
It could be that a clever manipulation of \eqref{eq:CCR} convexifies it just as we did for MEE. 

\vspace{-5pt}
\section{Conclusion}

Many problems in geometry processing suffer from nonlinearity, forcing users to cope with only locally optimal solutions. 
This nonlinearity makes it challenging to leave the realm of piecewise linear geometry, where at least low-level operations such as closest-point queries, injectivity testing, intersection, and collision detection can be solved with confidence, a solid base facilitating development of more complex algorithms. 

With our SOS framework, these and other low level operations are readily extended to higher-order surfaces. SOS geometry processing alleviates concerns, for example, that curved surface continuous collision detection might miss collisions due to linearization error, leading to unrealizable states. Similarly, SOS-based closest projection does not suffer from local optima that could affect processes down the line. 
SOS programming  transforms these and other problems on a huge variety of curved patch types into problems where a user can confidently certify a global optimum. 

SOS programming suggests a broadly-applicable framework for development of geometry processing algorithms.  To fill in the details of this new approach, \Cref{sec:discussion} presents a variety of concrete problems for future research at the intersection of SOS programming and geometry processing.
At a coarser level, our work serves as a stepping stone toward design of flexible simulation tools for curved surface representations. Recent work by \citet{Trusty_Chen_Levin_SEM_2021} also progresses in this direction but lacks modules for exactly resolving intersection and collision. Our tools enable confident collision detection operations analogous to the ones applied to piecewise-linear surfaces and can readily be incorporated into simulation algorithms. This opens the way toward collision response tools for a variety of geometric representations, including the velocity filter in \Cref{sec:alternatequant}. 

SOS geometry processing also supports interactions between multiple element types. While many geometry processing and simulation tools expect a unified set of, e.g., only triangle or hex elements, one can use SOS tools to formulate---for example---colliding a NURBS surface with a quadratic B\'ezier triangle mesh. Another interesting extension would be to NURBS patches with trimming curves, which could be implemented with more general domain constraint polynomials $g_i$ along with the denominator-clearing method from \Cref{sec:outside-template}. One only needs to describe the element types as polynomials to the SOS framework. Beyond developing numerical tools, a programming language-inspired approach to geometry processing might consider what level of abstraction a language needs to expose this flexibility to a user.

These opportunities for further development aside, our broad SOS framework and specific model problems are already beneficial to geometry processing. These problems would otherwise each require their own solutions, including heuristics for number of initial points, density of linearization, tolerance for intersection, gradient step size, and many other nonlinear optimization parameters, all of which must be tuned per surface type. SOS instead provides a single unified framework for common objectives across the most popular surface representations.

\vspace{-2.5pt}
\begin{acks}
The authors would like to thank Diego Cifuentes for much help understanding SOS theory and Misha Bessmeltsev, Etienne Vouga, Ilya Baran, and Paul Stallings for valuable suggestions and discussion. We would also like to thank Dmitriy Smirnov and Ty Trusty for providing several models used in the figures.

Paul Zhang acknowledges the generous support of the DOE Computational Science Graduate Fellowship. David Palmer acknowledges the generous support of the Hertz Fellowship and MathWorks Fellowship.

The MIT Geometric Data Processing group acknowledges the generous support of ARO grant W911NF2010168, of AFOSR award FA9550-19-1-031, of NSF grants IIS-1838071 and CHS-1955697, from the CSAIL Systems that Learn program, from the MIT–IBM Watson AI Laboratory, from the Toyota–CSAIL Joint Research Center, from a gift from Adobe Systems, from an MIT.nano Immersion Lab/NCSOFT Gaming Program seed grant, and from the Skoltech–MIT Next Generation Program. 
\end{acks}

\vspace{-2.5pt}
\bibliographystyle{ACM-Reference-Format}
\bibliography{sos-bibliography.bib}


\begin{thebibliography}{31}


\ifx \showCODEN    \undefined \def \showCODEN     #1{\unskip}     \fi
\ifx \showDOI      \undefined \def \showDOI       #1{#1}\fi
\ifx \showISBNx    \undefined \def \showISBNx     #1{\unskip}     \fi
\ifx \showISBNxiii \undefined \def \showISBNxiii  #1{\unskip}     \fi
\ifx \showISSN     \undefined \def \showISSN      #1{\unskip}     \fi
\ifx \showLCCN     \undefined \def \showLCCN      #1{\unskip}     \fi
\ifx \shownote     \undefined \def \shownote      #1{#1}          \fi
\ifx \showarticletitle \undefined \def \showarticletitle #1{#1}   \fi
\ifx \showURL      \undefined \def \showURL       {\relax}        \fi
\providecommand\bibfield[2]{#2}
\providecommand\bibinfo[2]{#2}
\providecommand\natexlab[1]{#1}
\providecommand\showeprint[2][]{arXiv:#2}

\bibitem[\protect\citeauthoryear{Ahmadi, Hall, Makadia, and Sindhwani}{Ahmadi
  et~al\mbox{.}}{2017}]%
        {Ahmadi2017sos}
\bibfield{author}{\bibinfo{person}{Amir Ahmadi}, \bibinfo{person}{Georgina
  Hall}, \bibinfo{person}{Ameesh Makadia}, {and} \bibinfo{person}{Vikas
  Sindhwani}.} \bibinfo{year}{2017}\natexlab{}.
\newblock \showarticletitle{Geometry of 3D Environments and Sum of Squares
  Polynomials}.
\newblock
\urldef\tempurl%
\url{https://doi.org/10.15607/RSS.2017.XIII.071}
\showDOI{\tempurl}


\bibitem[\protect\citeauthoryear{Alizadeh}{Alizadeh}{1995}]%
        {alizadeh1995interior}
\bibfield{author}{\bibinfo{person}{Farid Alizadeh}.}
  \bibinfo{year}{1995}\natexlab{}.
\newblock \showarticletitle{Interior point methods in semidefinite programming
  with applications to combinatorial optimization}.
\newblock \bibinfo{journal}{\emph{SIAM Journal on Optimization}}
  \bibinfo{volume}{5}, \bibinfo{number}{1} (\bibinfo{year}{1995}),
  \bibinfo{pages}{13--51}.
\newblock
\urldef\tempurl%
\url{https://doi.org/10.1137/0805002}
\showDOI{\tempurl}


\bibitem[\protect\citeauthoryear{Barnhill, Farin, Jordan, and Piper}{Barnhill
  et~al\mbox{.}}{1987}]%
        {barnhill1987surface}
\bibfield{author}{\bibinfo{person}{Robert~E. Barnhill}, \bibinfo{person}{Gerald
  Farin}, \bibinfo{person}{M. Jordan}, {and} \bibinfo{person}{Bruce~R. Piper}.}
  \bibinfo{year}{1987}\natexlab{}.
\newblock \showarticletitle{Surface/surface intersection}.
\newblock \bibinfo{journal}{\emph{Computer Aided Geometric Design}}
  \bibinfo{volume}{4}, \bibinfo{number}{1-2} (\bibinfo{year}{1987}),
  \bibinfo{pages}{3--16}.
\newblock
\urldef\tempurl%
\url{https://doi.org/10.1016/0167-8396(87)90020-3}
\showDOI{\tempurl}


\bibitem[\protect\citeauthoryear{Blekherman, Parrilo, and Thomas}{Blekherman
  et~al\mbox{.}}{2012}]%
        {blekherman2012semidefinite}
\bibfield{author}{\bibinfo{person}{Grigoriy Blekherman},
  \bibinfo{person}{Pablo~A. Parrilo}, {and} \bibinfo{person}{Rekha~R. Thomas}.}
  \bibinfo{year}{2012}\natexlab{}.
\newblock \bibinfo{booktitle}{\emph{Semidefinite Optimization and Convex
  Algebraic Geometry}}.
\newblock \bibinfo{publisher}{SIAM}.
\newblock
\urldef\tempurl%
\url{https://doi.org/10.1137/1.9781611972290}
\showDOI{\tempurl}


\bibitem[\protect\citeauthoryear{Boyd and Vandenberghe}{Boyd and
  Vandenberghe}{2004}]%
        {boyd2004convex}
\bibfield{author}{\bibinfo{person}{Stephen Boyd} {and} \bibinfo{person}{Lieven
  Vandenberghe}.} \bibinfo{year}{2004}\natexlab{}.
\newblock \bibinfo{booktitle}{\emph{Convex Optimization}}.
\newblock \bibinfo{publisher}{Cambridge University Press}.
\newblock
\urldef\tempurl%
\url{https://doi.org/10.1017/CBO9780511804441}
\showDOI{\tempurl}


\bibitem[\protect\citeauthoryear{Cardoze, Cunha, Miller, Phillips, and
  Walkington}{Cardoze et~al\mbox{.}}{2004}]%
        {cardoze2004bezier}
\bibfield{author}{\bibinfo{person}{David Cardoze}, \bibinfo{person}{Alexandre
  Cunha}, \bibinfo{person}{Gary~L. Miller}, \bibinfo{person}{Todd Phillips},
  {and} \bibinfo{person}{Noel Walkington}.} \bibinfo{year}{2004}\natexlab{}.
\newblock \showarticletitle{A bezier-based approach to unstructured moving
  meshes}. In \bibinfo{booktitle}{\emph{Proceedings of the twentieth annual
  symposium on Computational geometry}}. \bibinfo{pages}{310--319}.
\newblock
\urldef\tempurl%
\url{https://doi.org/10.1145/997817.997864}
\showDOI{\tempurl}


\bibitem[\protect\citeauthoryear{Catmull and Clark}{Catmull and Clark}{1978}]%
        {catmull1978recursively}
\bibfield{author}{\bibinfo{person}{Edwin Catmull} {and} \bibinfo{person}{James
  Clark}.} \bibinfo{year}{1978}\natexlab{}.
\newblock \showarticletitle{Recursively generated B-spline surfaces on
  arbitrary topological meshes}.
\newblock \bibinfo{journal}{\emph{Computer-aided design}} \bibinfo{volume}{10},
  \bibinfo{number}{6} (\bibinfo{year}{1978}), \bibinfo{pages}{350--355}.
\newblock
\urldef\tempurl%
\url{https://doi.org/10.1145/280811.280992}
\showDOI{\tempurl}


\bibitem[\protect\citeauthoryear{Jiang, Zhang, Hu, Schneider, Zorin, and
  Panozzo}{Jiang et~al\mbox{.}}{2021}]%
        {jiang2020highordertetmeshes}
\bibfield{author}{\bibinfo{person}{Zhongshi Jiang}, \bibinfo{person}{Ziyi
  Zhang}, \bibinfo{person}{Yixin Hu}, \bibinfo{person}{Teseo Schneider},
  \bibinfo{person}{Denis Zorin}, {and} \bibinfo{person}{Daniele Panozzo}.}
  \bibinfo{year}{2021}\natexlab{}.
\newblock \showarticletitle{Bijective and Coarse High-Order Tetrahedral
  Meshes}.
\newblock \bibinfo{journal}{\emph{ACM Transactions on Graphics}}
  (\bibinfo{year}{2021}).
\newblock
\urldef\tempurl%
\url{https://doi.org/10.1145/3450626.3459840}
\showDOI{\tempurl}


\bibitem[\protect\citeauthoryear{Kar{\v{c}}iauskas and
  Peters}{Kar{\v{c}}iauskas and Peters}{2020}]%
        {karvciauskas2020low}
\bibfield{author}{\bibinfo{person}{K{\k{e}}stutis Kar{\v{c}}iauskas} {and}
  \bibinfo{person}{J{\"o}rg Peters}.} \bibinfo{year}{2020}\natexlab{}.
\newblock \showarticletitle{Low degree splines for locally quad-dominant
  meshes}.
\newblock \bibinfo{journal}{\emph{Computer Aided Geometric Design}}
  \bibinfo{volume}{83} (\bibinfo{year}{2020}), \bibinfo{pages}{101934}.
\newblock
\urldef\tempurl%
\url{https://doi.org/10.1016/j.cagd.2020.101934}
\showDOI{\tempurl}


\bibitem[\protect\citeauthoryear{Koch, Matveev, Jiang, Williams, Artemov,
  Burnaev, Alexa, Zorin, and Panozzo}{Koch et~al\mbox{.}}{2019}]%
        {koch2019abc}
\bibfield{author}{\bibinfo{person}{Sebastian Koch}, \bibinfo{person}{Albert
  Matveev}, \bibinfo{person}{Zhongshi Jiang}, \bibinfo{person}{Francis
  Williams}, \bibinfo{person}{Alexey Artemov}, \bibinfo{person}{Evgeny
  Burnaev}, \bibinfo{person}{Marc Alexa}, \bibinfo{person}{Denis Zorin}, {and}
  \bibinfo{person}{Daniele Panozzo}.} \bibinfo{year}{2019}\natexlab{}.
\newblock \showarticletitle{{ABC}: A big cad model dataset for geometric deep
  learning}. In \bibinfo{booktitle}{\emph{Proceedings of the IEEE/CVF
  Conference on Computer Vision and Pattern Recognition}}.
  \bibinfo{pages}{9601--9611}.
\newblock


\bibitem[\protect\citeauthoryear{Lasserre}{Lasserre}{2001}]%
        {lasserre2001global}
\bibfield{author}{\bibinfo{person}{Jean~B. Lasserre}.}
  \bibinfo{year}{2001}\natexlab{}.
\newblock \showarticletitle{Global optimization with polynomials and the
  problem of moments}.
\newblock \bibinfo{journal}{\emph{SIAM Journal on Optimization}}
  \bibinfo{volume}{11}, \bibinfo{number}{3} (\bibinfo{year}{2001}),
  \bibinfo{pages}{796--817}.
\newblock
\urldef\tempurl%
\url{https://doi.org/10.1137/S1052623400366802}
\showDOI{\tempurl}


\bibitem[\protect\citeauthoryear{L{\"{o}}fberg}{L{\"{o}}fberg}{2004}]%
        {Lofberg2004}
\bibfield{author}{\bibinfo{person}{Johan L{\"{o}}fberg}.}
  \bibinfo{year}{2004}\natexlab{}.
\newblock \showarticletitle{{YALMIP}: A Toolbox for Modeling and Optimization
  in {MATLAB}}. In \bibinfo{booktitle}{\emph{In Proceedings of the CACSD
  Conference}}. \bibinfo{address}{Taipei, Taiwan}.
\newblock
\urldef\tempurl%
\url{https://doi.org/10.1109/CACSD.2004.1393890}
\showDOI{\tempurl}


\bibitem[\protect\citeauthoryear{Loop}{Loop}{1987}]%
        {loop1987smooth}
\bibfield{author}{\bibinfo{person}{Charles Loop}.}
  \bibinfo{year}{1987}\natexlab{}.
\newblock \showarticletitle{Smooth subdivision surfaces based on triangles}.
\newblock \bibinfo{journal}{\emph{Master's thesis, University of Utah,
  Department of Mathematics}} (\bibinfo{year}{1987}).
\newblock


\bibitem[\protect\citeauthoryear{Mandad and Campen}{Mandad and Campen}{2020}]%
        {mandad2020bezier}
\bibfield{author}{\bibinfo{person}{Manish Mandad} {and} \bibinfo{person}{Marcel
  Campen}.} \bibinfo{year}{2020}\natexlab{}.
\newblock \showarticletitle{B{\'e}zier guarding: precise higher-order meshing
  of curved 2D domains}.
\newblock \bibinfo{journal}{\emph{ACM Transactions on Graphics (TOG)}}
  \bibinfo{volume}{39}, \bibinfo{number}{4} (\bibinfo{year}{2020}),
  \bibinfo{pages}{103--1}.
\newblock
\urldef\tempurl%
\url{https://doi.org/10.1145/3386569.3392372}
\showDOI{\tempurl}


\bibitem[\protect\citeauthoryear{Marschner, Palmer, Zhang, and
  Solomon}{Marschner et~al\mbox{.}}{2020}]%
        {marschner2020hexahedral}
\bibfield{author}{\bibinfo{person}{Zo{\"e} Marschner}, \bibinfo{person}{David
  Palmer}, \bibinfo{person}{Paul Zhang}, {and} \bibinfo{person}{Justin
  Solomon}.} \bibinfo{year}{2020}\natexlab{}.
\newblock \showarticletitle{Hexahedral Mesh Repair via Sum-of-Squares
  Relaxation}. In \bibinfo{booktitle}{\emph{Computer Graphics Forum}},
  Vol.~\bibinfo{volume}{39}. Wiley Online Library, \bibinfo{pages}{133--147}.
\newblock
\urldef\tempurl%
\url{https://doi.org/doi.org/10.1111/cgf.14074}
\showDOI{\tempurl}


\bibitem[\protect\citeauthoryear{{MOSEK~ApS}}{{MOSEK~ApS}}{2017}]%
        {aps_mosek_2017}
\bibfield{author}{\bibinfo{person}{{MOSEK~ApS}}.}
  \bibinfo{year}{2017}\natexlab{}.
\newblock \bibinfo{booktitle}{\emph{The {MOSEK} optimization toolbox for
  {MATLAB} manual. {Version} 8.1.}}
\newblock
\urldef\tempurl%
\url{http://docs.mosek.com/8.1/toolbox/index.html}
\showURL{%
\tempurl}


\bibitem[\protect\citeauthoryear{Nesterov and Nemirovskii}{Nesterov and
  Nemirovskii}{1994}]%
        {nesterov-nemirovskii}
\bibfield{author}{\bibinfo{person}{Yurii Nesterov} {and}
  \bibinfo{person}{Arkadii Nemirovskii}.} \bibinfo{year}{1994}\natexlab{}.
\newblock \bibinfo{booktitle}{\emph{Interior-Point Polynomial Algorithms in
  Convex Programming}}.
\newblock \bibinfo{publisher}{Society for Industrial and Applied Mathematics}.
\newblock
\urldef\tempurl%
\url{https://doi.org/10.1137/1.9781611970791}
\showDOI{\tempurl}


\bibitem[\protect\citeauthoryear{Oh, Kim, Lee, Kim, and Elber}{Oh
  et~al\mbox{.}}{2012}]%
        {oh2012efficient}
\bibfield{author}{\bibinfo{person}{Young-Taek Oh}, \bibinfo{person}{Yong-Joon
  Kim}, \bibinfo{person}{Jieun Lee}, \bibinfo{person}{Myung-Soo Kim}, {and}
  \bibinfo{person}{Gershon Elber}.} \bibinfo{year}{2012}\natexlab{}.
\newblock \showarticletitle{Efficient point-projection to freeform curves and
  surfaces}.
\newblock \bibinfo{journal}{\emph{Computer Aided Geometric Design}}
  \bibinfo{volume}{29}, \bibinfo{number}{5} (\bibinfo{year}{2012}),
  \bibinfo{pages}{242--254}.
\newblock
\urldef\tempurl%
\url{https://doi.org/10.1016/j.cagd.2011.04.002}
\showDOI{\tempurl}


\bibitem[\protect\citeauthoryear{Parillo}{Parillo}{2019}]%
        {ParilloClass}
\bibfield{author}{\bibinfo{person}{Pablo~A. Parillo}.}
  \bibinfo{year}{2019}\natexlab{}.
\newblock \bibinfo{title}{Algebraic Techniques and Semidefinite Optimization}.
\newblock
  \bibinfo{howpublished}{\url{https://learning-modules.mit.edu/materials/index.html?uuid=/course/6/sp19/6.256\#materials}}.
\newblock


\bibitem[\protect\citeauthoryear{Pekerman, Elber, and Kim}{Pekerman
  et~al\mbox{.}}{2008}]%
        {pekerman2008self}
\bibfield{author}{\bibinfo{person}{Diana Pekerman}, \bibinfo{person}{Gershon
  Elber}, {and} \bibinfo{person}{Myung-Soo Kim}.}
  \bibinfo{year}{2008}\natexlab{}.
\newblock \showarticletitle{Self-intersection detection and elimination in
  freeform curves and surfaces}.
\newblock \bibinfo{journal}{\emph{Computer-Aided Design}} \bibinfo{volume}{40},
  \bibinfo{number}{2} (\bibinfo{year}{2008}), \bibinfo{pages}{150--159}.
\newblock
\urldef\tempurl%
\url{https://doi.org/10.1016/j.cad.2007.10.004}
\showDOI{\tempurl}


\bibitem[\protect\citeauthoryear{Prajna, Papachristodoulou, and Parrilo}{Prajna
  et~al\mbox{.}}{2002}]%
        {prajna2002introducing}
\bibfield{author}{\bibinfo{person}{Stephen Prajna}, \bibinfo{person}{Antonis
  Papachristodoulou}, {and} \bibinfo{person}{Pablo~A. Parrilo}.}
  \bibinfo{year}{2002}\natexlab{}.
\newblock \showarticletitle{Introducing SOSTOOLS: A general purpose sum of
  squares programming solver}. In \bibinfo{booktitle}{\emph{Proceedings of the
  41st IEEE Conference on Decision and Control, 2002.}},
  Vol.~\bibinfo{volume}{1}. IEEE, \bibinfo{pages}{741--746}.
\newblock


\bibitem[\protect\citeauthoryear{Putinar}{Putinar}{1993}]%
        {putinar1993positive}
\bibfield{author}{\bibinfo{person}{Mihai Putinar}.}
  \bibinfo{year}{1993}\natexlab{}.
\newblock \showarticletitle{Positive polynomials on compact semi-algebraic
  sets}.
\newblock \bibinfo{journal}{\emph{Indiana University Mathematics Journal}}
  \bibinfo{volume}{42}, \bibinfo{number}{3} (\bibinfo{year}{1993}),
  \bibinfo{pages}{969--984}.
\newblock


\bibitem[\protect\citeauthoryear{Sawhney and Crane}{Sawhney and Crane}{2020}]%
        {Sawhney:2020:MCG}
\bibfield{author}{\bibinfo{person}{Rohan Sawhney} {and} \bibinfo{person}{Keenan
  Crane}.} \bibinfo{year}{2020}\natexlab{}.
\newblock \showarticletitle{Monte Carlo Geometry Processing: A Grid-Free
  Approach to PDE-Based Methods on Volumetric Domains}.
\newblock \bibinfo{journal}{\emph{ACM Trans. Graph.}} \bibinfo{volume}{39},
  \bibinfo{number}{4} (\bibinfo{year}{2020}).
\newblock
\urldef\tempurl%
\url{https://doi.org/10.1145/3386569.3392374}
\showDOI{\tempurl}


\bibitem[\protect\citeauthoryear{Schneider, Dumas, Gao, Botsch, Panozzo, and
  Zorin}{Schneider et~al\mbox{.}}{2019}]%
        {Schneider:2019:PFM}
\bibfield{author}{\bibinfo{person}{Teseo Schneider},
  \bibinfo{person}{J{\'e}r{\'e}mie Dumas}, \bibinfo{person}{Xifeng Gao},
  \bibinfo{person}{Mario Botsch}, \bibinfo{person}{Daniele Panozzo}, {and}
  \bibinfo{person}{Denis Zorin}.} \bibinfo{year}{2019}\natexlab{}.
\newblock \showarticletitle{Poly-Spline Finite-Element Method}.
\newblock \bibinfo{journal}{\emph{ACM Trans. Graph.}} \bibinfo{volume}{38},
  \bibinfo{number}{3}, Article \bibinfo{articleno}{19} (\bibinfo{date}{March}
  \bibinfo{year}{2019}), \bibinfo{numpages}{16}~pages.
\newblock
\showISSN{0730-0301}
\urldef\tempurl%
\url{https://doi.org/10.1145/3313797}
\showDOI{\tempurl}


\bibitem[\protect\citeauthoryear{Schneider, Hu, Dumas, Gao, Panozzo, and
  Zorin}{Schneider et~al\mbox{.}}{2018}]%
        {Schneider:2018:DSA}
\bibfield{author}{\bibinfo{person}{Teseo Schneider}, \bibinfo{person}{Yixin
  Hu}, \bibinfo{person}{Jérémie Dumas}, \bibinfo{person}{Xifeng Gao},
  \bibinfo{person}{Daniele Panozzo}, {and} \bibinfo{person}{Denis Zorin}.}
  \bibinfo{year}{2018}\natexlab{}.
\newblock \showarticletitle{Decoupling Simulation Accuracy from Mesh Quality}.
\newblock \bibinfo{journal}{\emph{ACM Transactions on Graphics}}
  \bibinfo{volume}{37}, \bibinfo{number}{6} (\bibinfo{date}{10}
  \bibinfo{year}{2018}).
\newblock
\urldef\tempurl%
\url{https://doi.org/10.1145/3272127.3275067}
\showDOI{\tempurl}


\bibitem[\protect\citeauthoryear{Smirnov, Bessmeltsev, and Solomon}{Smirnov
  et~al\mbox{.}}{2020a}]%
        {smirnov2020learning}
\bibfield{author}{\bibinfo{person}{Dmitriy Smirnov}, \bibinfo{person}{Mikhail
  Bessmeltsev}, {and} \bibinfo{person}{Justin Solomon}.}
  \bibinfo{year}{2020}\natexlab{a}.
\newblock \showarticletitle{Learning Manifold Patch-Based Representations of
  Man-Made Shapes}, In \bibinfo{booktitle}{International Conference on Learning
  Representations}.
\newblock \bibinfo{journal}{\emph{arXiv preprint arXiv:1906.12337}}.
\newblock


\bibitem[\protect\citeauthoryear{Smirnov, Fisher, Kim, Zhang, and
  Solomon}{Smirnov et~al\mbox{.}}{2020b}]%
        {smirnov2020deep}
\bibfield{author}{\bibinfo{person}{Dmitriy Smirnov}, \bibinfo{person}{Matthew
  Fisher}, \bibinfo{person}{Vladimir~G. Kim}, \bibinfo{person}{Richard Zhang},
  {and} \bibinfo{person}{Justin Solomon}.} \bibinfo{year}{2020}\natexlab{b}.
\newblock \showarticletitle{Deep parametric shape predictions using distance
  fields}. In \bibinfo{booktitle}{\emph{Proceedings of the IEEE/CVF Conference
  on Computer Vision and Pattern Recognition}}. \bibinfo{pages}{561--570}.
\newblock
\urldef\tempurl%
\url{https://doi.org/10.1109/CVPR42600.2020.00064}
\showDOI{\tempurl}


\bibitem[\protect\citeauthoryear{Todd}{Todd}{2016}]%
        {todd2016minimum}
\bibfield{author}{\bibinfo{person}{Michael~J Todd}.}
  \bibinfo{year}{2016}\natexlab{}.
\newblock \bibinfo{booktitle}{\emph{Minimum-volume ellipsoids: Theory and
  algorithms}}.
\newblock \bibinfo{publisher}{SIAM}.
\newblock
\urldef\tempurl%
\url{https://doi.org/10.1137/1.9781611974386}
\showDOI{\tempurl}


\bibitem[\protect\citeauthoryear{Toh, Todd, and T\"{u}t\"{u}nc\"{u}}{Toh
  et~al\mbox{.}}{2001}]%
        {tutuncu2001sdpt3}
\bibfield{author}{\bibinfo{person}{Kim-Chuan Toh}, \bibinfo{person}{Michael~J.
  Todd}, {and} \bibinfo{person}{Reha~H. T\"{u}t\"{u}nc\"{u}}.}
  \bibinfo{year}{2001}\natexlab{}.
\newblock \showarticletitle{SDPT3—a Matlab software package for
  semidefinite-quadratic-linear programming, version 3.0}.
\newblock \bibinfo{journal}{\emph{Web page http://www. math. nus. edu.
  sg/mattohkc/sdpt3. html}} (\bibinfo{year}{2001}).
\newblock
\urldef\tempurl%
\url{https://doi.org/10.1080/10556789908805762}
\showDOI{\tempurl}


\bibitem[\protect\citeauthoryear{Trusty, Chen, and Levin}{Trusty
  et~al\mbox{.}}{2021}]%
        {Trusty_Chen_Levin_SEM_2021}
\bibfield{author}{\bibinfo{person}{Ty Trusty}, \bibinfo{person}{Honglin Chen},
  {and} \bibinfo{person}{David~I.W. Levin}.} \bibinfo{year}{2021}\natexlab{}.
\newblock \showarticletitle{The Shape Matching Element Method: Direct Animation
  of Curved Surface Models}.
\newblock \bibinfo{journal}{\emph{ACM Transactions on Graphics}}
  (\bibinfo{year}{2021}).
\newblock
\urldef\tempurl%
\url{https://doi.org/10.1145/3450626.3459772}
\showDOI{\tempurl}


\bibitem[\protect\citeauthoryear{Wawrzinek, Hildebrandt, and
  Polthier}{Wawrzinek et~al\mbox{.}}{2011}]%
        {PE:VMV:VMV11:113-120}
\bibfield{author}{\bibinfo{person}{Anna Wawrzinek}, \bibinfo{person}{Klaus
  Hildebrandt}, {and} \bibinfo{person}{Konrad Polthier}.}
  \bibinfo{year}{2011}\natexlab{}.
\newblock \showarticletitle{{Koiter's Thin Shells on Catmull-Clark Limit
  Surfaces}}. In \bibinfo{booktitle}{\emph{Vision, Modeling, and Visualization
  (2011)}}, \bibfield{editor}{\bibinfo{person}{Peter Eisert},
  \bibinfo{person}{Joachim Hornegger}, {and} \bibinfo{person}{Konrad Polthier}}
  (Eds.). \bibinfo{publisher}{The Eurographics Association}.
\newblock
\showISBNx{978-3-905673-85-2}
\urldef\tempurl%
\url{https://doi.org/10.2312/PE/VMV/VMV11/113-120}
\showDOI{\tempurl}


\end{thebibliography}
\end{document}